\documentclass[pdflatex,sn-mathphys]{sn-jnl}% Math and Physical Sciences Reference Style
\usepackage{comment}
\usepackage{amsmath}
\usepackage{amssymb}
\usepackage{amsfonts}
\usepackage{soul}
\theoremstyle{thmstyleone}%
\newcommand{\avg}[1]{\langle #1 \rangle}
\theoremstyle{thmstyletwo}%

\theoremstyle{thmstylethree}%

\begin{document}

\title[Characterising the dynamics of unlabelled temporal networks]{Characterising the dynamics of unlabelled temporal networks}

\author{Annalisa Caligiuri\footnote{Corresponding author. Email: \url{annalisa@ifisc.uib-csic.es.}}$^{\ 1}$, Tobias Galla$^{1}$ and Lucas Lacasa$^{1}$}

\affil[1]{Institute for Cross-Disciplinary Physics and Complex Systems (IFISC, CSIC-UIB), 07122 Palma de Mallorca (Spain)}

\abstract{
Networks model the architecture backbone of complex systems. The backbone itself can change over time leading to what is called `temporal networks'. Interpreting temporal networks as trajectories in graph space of a latent graph dynamics has recently enabled the extension of concepts and tools from dynamical systems and time series to networks. Here we address temporal networks with unlabelled nodes, a case that has received relatively little attention so far. Situations in which node labelling cannot be tracked over time often emerge in practice due to technical challenges, or privacy constraints. In unlabelled temporal networks there is no one-to-one matching between a network snapshot and its adjacency matrix. Characterizing the dynamical properties of such unlabelled network trajectories is nontrivial. We here exploit graph invariants to extend to the unlabelled setting network-dynamical quantifiers of linear correlations and dynamical instability. In particular, we focus on autocorrelation functions and the sensitive dependence on initial conditions. We show with synthetic graph dynamics that the measures are capable of recovering and estimating these dynamical fingerprints even when node labels are  unavailable. We also validate the methods for some empirical temporal networks with removed node labels.
}


%%\pacs[JEL Classification]{D8, H51}

%%\pacs[MSC Classification]{35A01, 65L10, 65L12, 65L20, 65L70}
\maketitle

\vspace{5em}

\section{Introduction}\label{Sec:introduction}
The architecture of many natural and artificial systems --ranging from social interactions \cite{zhao2011social,palla2009social} to the brain \cite{bassett2017network,dimitriadis2010tracking}-- can be modeled as networks or graphs. The field of Network Science \cite{latora2017complex, newman2018networks} provides insights into the effects the structure of interactions has on the dynamics of a system. Temporal networks (also called time-varying graphs) \cite{HOLME201297,masuda2016guide,Nicosia2013} describe situations in which the interaction between elements itself changes in time. The use of temporal --instead of static-- network models has been instrumental to unveiling more nuanced collective phenomena, from diffusion \cite{masuda2013temporal, delvenne2015diffusion, scholtes2014causality}, social \cite{Starnini:2013} or financial interactions \cite{mazzarisi2020dynamic} to epidemic spreading \cite{hiraoka2018correlated, van2013non}, brain activity \cite{thompson2017static} or even propagation of delays in the air transport system \cite{zanin2009dynamics, zanin2013modelling}. 
More recently, the intrinsic dynamics {\it of} temporal networks has started to attract attention, and one focus is on the characterisation of these dynamics with the tools of dynamical systems, time series and signal processing \cite{Williams_2019, williams2022shape, lacasa2022correlations, caligiuri2023, danovski2024dynamical, thongprayoon2023embedding, hartle2024autocorrelation, lacasa2024scalar}.

\medskip \noindent 
Among other definitions, a temporal network can be introduced as an ordered sequence of network {\it snapshots} ${\cal S}_{\cal G}=(G_1,G_2,\dots,G_T)$, where the $t$-snapshot $G_t$ characterises the interaction architecture of the elements of the system at the $t$-th discrete time window. In other words, each network snapshot is a graph that aggregates the activity in the system within the corresponding time window, and thus all interactions which are present at some point within such time window are annotated as links between nodes in such graph. Accordingly, $\cal S_G$ is also a time series of graphs, and thus can be interpreted as a network trajectory in graph space \cite{lacasa2022correlations}, i.e. the observed trajectory of a (latent) graph dynamical system.

\medskip \noindent 
In many applications, the nodes of the network snapshots are known and labelled, and there is a clear tracking of which node is which throughout each snapshot of the network trajectory. These graphs are referred to as {\it labelled} networks. Each snapshot $G_t$ is then fully characterized by its adjacency matrix $A_t$, and thus ${\cal S}_{\cal G}$ can equivalently be described by the sequence of adjacency matrices ${\cal S}_{\cal G}\equiv {\cal S}_{\cal A}:= (A_1,A_2,\dots, A_T)$. 
As a matter of fact, most work on temporal networks explicitly or implicitly assumes that network snapshots are labelled, and indeed this is convenient as it enables the use of powerful methods from linear algebra for the analysis of these networks. However, in real-world applications this assumption often requires sophisticated methods to accurately track elements of a system --the nodes-- over time. There are indeed cases in which matching labels between snapshots of a temporal network is either technically challenging or outright impossible: examples include proximity networks of animal or biological swarms \cite{Attanasi2013GReTAANG, AnimalTracking}, or social interactions subject to privacy constraints, e.g. in epidemic spreading \cite{zhang2010privacy,zheleva2011privacy,cencetti2021digital, rodriguez2021population}, to cite a few. In other cases, the constraint is more of an inherent property of the system: for instance, in chemistry sometimes we need to compare the chemical structure of two different drugs. It is not obvious how to match nodes for graph comparison of these molecular graphs, something called the molecular similarity problem \cite{lopez2024molecular}. Finally, sometimes nodes themselves are genuinely indistinguishable (in the sense of quantum physics).\\
If nodes are indistinguishable no meaningful adjacency matrix can be assigned. But even in cases where the ground truth labels exist (but are hidden, or such information is not available), network snapshots cannot be uniquely represented by an adjacency matrix and one is left with a sequence of unlabelled networks. As a matter of fact, an unlabelled graph of $N$ nodes does not have a unique adjacency matrix, but a whole set of $N!$ equivalent matrix representations: the graph's so-called automorphism group \cite{bondy2008graph}. These different representations are generated from one another through row-column permutations in the matrix, i.e. permutation of the node set. While one can always assign a valid node labelling at random to each snapshot and thus construct a sequence of adjacency matrices ${\cal S}_{\cal A}$ from ${\cal S}_{\cal G}$ such a procedure is problematic, as it removes any possible structure in the dynamics of the original sequence of networks. Thus in general one cannot use a sequence of randomly assigned node labels to assess the dynamics of unlabelled network trajectories.

\medskip \noindent 
Many important problems in the mathematical field of graph theory are related to unlabelled graphs, including the celebrated graph isomorphism problem \cite{babai1980,Yan2016,mckay2014practical}. This is the question how to efficiently decide if two graphs are isomorphic (i.e. whether their unlabelled structure is the same, and thus whether there exists a row-column permutation to go from the adjacency matrix of one graph to the other). A more general question is how to quantitatively compare unlabelled graphs, i.e., to say how similar or dissimilar two unlabelled graphs are. For labelled graphs this question can be readily assessed by comparing their adjacency matrices via any kind of matrix norm, but in the case of unlabelled graphs, such a comparison is more tricky. Various possibilities exist, from graph canonization \cite{canonization} to the definition of several pseudo-distances \cite{practitioner_guide, schieber2017quantification} --each of them based on different graph properties--. Nonetheless, there is to date no computationally efficient one-size-fits-all solution (note that this would indirectly solve the graph isomorphism problem).

\medskip \noindent 
Aside from classic graph-theoretical questions \cite{wright1976evolution, bollobas1982asymptotic, walsh1978k}, complexity-oriented research on unlabelled networks has been scarce to date, with the notable exceptions of works dealing with entropic and other aspects of unlabelled graph ensembles \cite{hartle2020network, paton2022entropy}.
In this paper, we address the problem of characterizing the dynamical properties of unlabelled network trajectories. Our general philosophy is to use examples of synthetic or real-world labelled temporal networks as ground-truth starting point. We then remove node labels, resulting in an unlabelled sequence of graphs. Subsequently, we then ask if and how dynamic properties of the original labelled sequence can be recovered from the time series of unlabelled graphs. 
We circumvent the problem of the lack of labels by focusing on graph invariants. These are graph properties that remain invariant under row-column permutation and therefore are the same for all adjacency matrices within the automorphism group of an unlabelled graph. As a proof of concept, we use three graph invariants to characterize different types of graph dissimilarities. We subsequently build on these measures to characterize, in the unlabelled setting, dynamical properties such as collective network periodicity (i.e. network pulsation) and linear temporal correlations, and their estimation via a suitable (unlabelled) network autocorrelation function \cite{lacasa2022correlations}. Additionally, we also investigate the problem of characterizing the stability or instability of orbits, and sensitivity to initial conditions in unlabelled network trajectories \cite{caligiuri2023, danovski2024dynamical}. Our results show that the characterization of such dynamical properties is possible, and that there are  important differences compared to the labelled case.

\medskip \noindent 
The rest of the paper is organized as follows: In Sec.~\ref{Sec:Distances}, we introduce three permutation-invariant pseudo-distances that will be used throughout the paper to subsequently build the dynamical quantifiers. We analyze their behavior and performance as dissimilarity measures. Section~\ref{Sec:metrics} then explores the unlabelled version of the network autocorrelation function (which quantifies periodicity and memory in unlabelled network trajectories) and the problem of measuring sensitivity to initial conditions in chaotic (unlabelled) network trajectories. We validate these in a range of synthetic processes, encompassing models of periodic and chaotic network dynamics, as well as models with nontrivial memory, and apply the method to some empirical temporal networks with hidden label information. Finally, in Sec.~\ref{Sec:conclusion} we conclude and discuss potential research avenues arising from this work.

\section{Comparing unlabelled graphs}\label{Sec:Distances}
\subsection{Pseudo-distances from graph invariants}
The analysis of unlabelled network trajectories requires one to be able to compare unlabelled graphs. Let $G$ and $H$ be two generic unlabelled graphs. It is not straightforward to use any notion of matrix distance to assess how similar $G$ and $H$ are, as an unlabelled graph does not have a unique adjacency matrix. Indeed, if adjacency matrices are constructed by introducing node labels, the matrix distance between two unlabelled graphs may depend on the choice of node labels and not be an intrinsic property of the graphs.\\
A partial way out is to not use any adjacency matrix as the representation of the graphs $G$ and $H$, but to represent the graphs via some (computationally efficient) properties which are invariant under node relabelling (i.e., properties that remain invariant under row/column permutation of the adjacency matrix). 
We highlight at this point that these graph invariants will not, in general, capture the full information of the graph: one can always find two graphs which are not isomorphic but have the same value of a given graph invariant. Nonetheless, the idea is that to a large extent these graph invariants still capture valuable information of  graphs, and that we can use this information to build a notion of graph dissimilarity.\\
Simple graph invariants which can be obtained in an efficient manner computationally include the following: the degree sequence (and statistical properties derived from it, such as the average degree, the degree distribution, and so on), centrality sequences, the spectrum of the adjacency matrix or the Laplacian matrix, and many others. Below we will construct dissimilarity measures between unlabelled graphs based on some of these properties. More precisely, we will use these invariants to construct {\it permutation-invariant pseudo-distances} between unlabelled graphs. We stress that distinct graphs can have the same value of simple graph invariants, resulting in a vanishing pseudo-distance. For two given graphs $G$ and $H$, we use the notation $d(G, H)$, often supplemented by a subscript to indicate the exact type of pseudo-distance.

\medskip \noindent 
Of course, the choice of the graph invariant is important, and in a particular application this choice can be informed  by the type of graph dynamics at hand. For instance, the total number of edges in a graph is a very simple graph invariant. While this is an important property of the graph, defining a pseudo-distance between two graphs e.g. as the absolute difference in the number of edges will lack expressivity for graph dynamics in which the total number of edges is preserved.\\

\subsection{Three concrete examples of permutation-invariant pseudo-distances}
We illustrate the use of permutation-invariant pseudo-distances by choosing three graph invariants: the degree sequence, the spectrum of the adjacency matrix, and the eigenvector centrality vector. In making this construction, we require the graphs $G$ and $H$ to have the same number of nodes (relaxing this constraint is left for future work), but otherwise they can have arbitrary edge sets and, more importantly, these pseudo-distances can be applied to both labelled and unlabelled graphs. {We always work with simple graphs, meaning there are no multiple edges between two nodes and no self-loops.}
\\

\noindent \textbf{Degree sequence pseudo-distance ($\mathbf{d_{deg}}$)} --
Given two graphs $G$ and $H$, each with $N$ nodes, we define the degree sequence-based pseudo-distance $d_{\text{deg}}(G,H)$ as a (properly normalized) $L_1$ distance between the degree sequences:
\begin{equation}\label{Eq:DSD}
    d_{\text{deg}}(G, H) = \frac{1}{N(N-1)}\sum_{i=1}^{N} \vert k_{i}^{G} - k_{i}^{H} \vert
\end{equation}
where $k_{i}^{G}$ and $k_{i}^{H}$ are the degrees occupying the $i$-th position of the ordered degree sequence of graphs $G$ and $H$, respectively.  We recall that the degree sequence of a graph is the sequence of the degrees of all nodes, usually sorted in non-increasing order. It encodes information on the graph's connectivity and its heterogeneity. The $k_i^H$ and $k_i^G$ take values between zero and $N-1$, hence the term $1/[N(N-1)]$ in Eq.~(\ref{Eq:DSD}) is a normalizing factor ensuring that $d_{\text{deg}}(G, H)\in [0,1]$.\\
The metric $d_{\text{deg}}(G, H)$ is  useful for graph dynamics for which the degree sequence changes in time. Since the degree sequence remains invariant under row-column permutations of the adjacency matrix, it is therefore identical for isomorphic graphs. The converse is not necessarily true (two graphs can have the same degree sequence but not be isomorphic to one another), hence while $d_{\text{deg}}(\cdot,\cdot)$ is a distance between degree sequences, it is only a pseudo-distance between graphs.\\

\noindent \textbf{Spectral pseudo-distance ($\mathbf{d_{spec}}$)} -- This is the $L_1$ distance between the spectra of the adjacency matrices of two graphs \cite{JOVANOVIC20121425}:

\begin{equation}\label{Eq:SD}
    d_{\text{spec}}(G,H)=\frac{ 1}{2N(N-1)} \sum^N_i \vert \lambda^{A_G}_i-\lambda^{A_H}_i \vert,
\end{equation}
where $\lambda^{A_G}_i$ and $\lambda^{A_H}_i$ are the $i$-th eigenvalues (once they have been ordered) of the adjacency matrices of graphs $G$ and $H$, respectively. The spectrum of a matrix remains unchanged under row-column permutations and is therefore the same for isomorphic graphs (again, while isomorphic graphs are isospectral, the converse is not necessarily true, hence $d_{\text{spec}}(\cdot,\cdot)$ is pseudo-distance between graphs).
The normalization in Eq.~(\ref{Eq:SD}) accounts for the fact that the maximum eigenvalue of the ajdacency matrix of a simple graph is always less than or equal to $N-1$ \cite{brouwer2011spectra}. Additionally, the smallest eigenvalue is greater than or equal to $1-N$. Therefore, the largest possible difference between two eigenvalues is $2(N-1)$.\\

\noindent \textbf{Eigenvector centrality pseudo-distance ($\mathbf{d_{eig}}$)} -- We note that $d_{\text{deg}}$ uses the degree sequence of the network, also called the degree centrality vector $(k_1,\dots,k_N$), ordered according to size. The same idea can be extended to centrality properties other than the degrees of the nodes. For illustration, we now consider the so-called eigenvector centrality, i.e. we rank the nodes' importance based on the eigenvector associated with the largest eigenvalue of the adjacency matrix \cite{BONACICH2007555,Schoenberg1969}. Let $\mathbf{v}^{G}=(v_1^G,v_2^G,\dots,v_N^G)$ and $\mathbf{v}^{H}=(v_1^H,v_2^H,\dots,v_N^H)$ be the  eigenvector centrality vectors of graphs $G$ and $H$. We assume that the components of these vectors are again ordered (i.e. we sort the vectors). We also assume that the vectors are normalised, i.e., $\sqrt{\sum_i (v_i^G)^2}=1$ and similarly for $\mathbf{v}^H$. Then, we define $d_{\text{eig}}$ as the $L_2$ distance between the eigenvector centrality vectors:
\begin{equation}\label{Eq:EC}
    d_{\text{eig}}(G,H)=\frac{ 1}{2} \sqrt{\sum^N_{i=1}  \left( v^{G}_i-v^{H}_i \right)^2}.
\end{equation}
The prefactor $1/2$ ensures that $d_\text{eig}$ remains in the interval from zero to one (Because of the normalisation of $\mathbf{v}^G$ and $\mathbf{v}^H$, we have $2 = \| \mathbf{v}^G \| + \| \mathbf{v}^H \| \geq \| \mathbf{v}^G - \mathbf{v}^H\| $
using the triangle inequality, and where $\vert \vert \cdot \vert \vert $ is the Euclidean norm). Similarly as before, the sorted eigenvectors remain unchanged under row-column permutations but there exist non-isomorphic graphs with the same eigenvector centrality, hence $d_{\text{eig}}(\cdot,\cdot)$ is a pseudo-distance between graphs.

\subsection{Exploring the behavior of $d_\text{deg}$, $d_\text{spec}$ and $d_\text{eig}$}

Before leveraging the graph invariants --and the resulting pseudo-distances-- in the task of analysing the dynamics of unlabelled network trajectories, we need to properly assess their behavior in quantifying dissimiliarities between pairs of graphs $G$ and $H$ in some controlled setting. To do that, in Secs.~\ref{sec:er} and \ref{sec:displace} we will define two different models that generate labelled graphs $G$ and $H$ such that we can tune how different they are based on a certain control parameter. This parameter  intuitively measures the size of the perturbation performed in $G$ to create $H$. To have a ground true distance between $G$ and $H$ we use a matrix norm, i.e. we measure the distance between the labelled graphs. We do this by using a normalized version of the element-wise $L_1$-distance between the adjacency matrices of $G$ and $H$,
\begin{equation}\label{eq:lab}
    d_{\text{lab}}(G,H) =\frac{1}{{\cal Z}}\sum_{i,j=1}^N \vert a_{ij} - b_{ij}\vert.
\end{equation}
 Here ${\cal Z}$ is a normalization factor which will be chosen differently in the context of the two models below to ensure that $d_\text{lab}$ remains between zero and one (details will be given below). The $a_{ij}$ and $b_{ij}$ are the $ij$-entries of the adjacency matrices of $G$ and $H$, respectively. Again, we assume that $G$ and $H$ have the same number of nodes.\\
 We will first prove that $d_{\text{lab}}(G,H)$ is actually linearly related to the parameter that controls the perturbation imposed in $G$ to create $H$, and that, accordingly, we can use $d_{\text{lab}}(G,H)$ as the ground truth quantity. Subsequently, we will proceed to remove all information on the node labels in $G$ and $H$, and will systematically compare all three unlabelled pseudo-distances $d_X(G,H)$ (where $X$ stands for `deg', `spec', or `eig') to the ground truth quantity, i.e., to $d_{\text{lab}}$.
 In particular, we are interested in assessing whether $d_X(G,H)$ is a non-decreasing function of $d_{\text{lab}}$, whether it shows a linear dependence or non-linear monotonic dependence, and if there are discontinuities. These properties will become important later when we use the pseudo-distances to quantify the dynamical behavior of unlabelled network trajectories.

\begin{figure}[htb]
\centering
\includegraphics[width=0.49\linewidth]{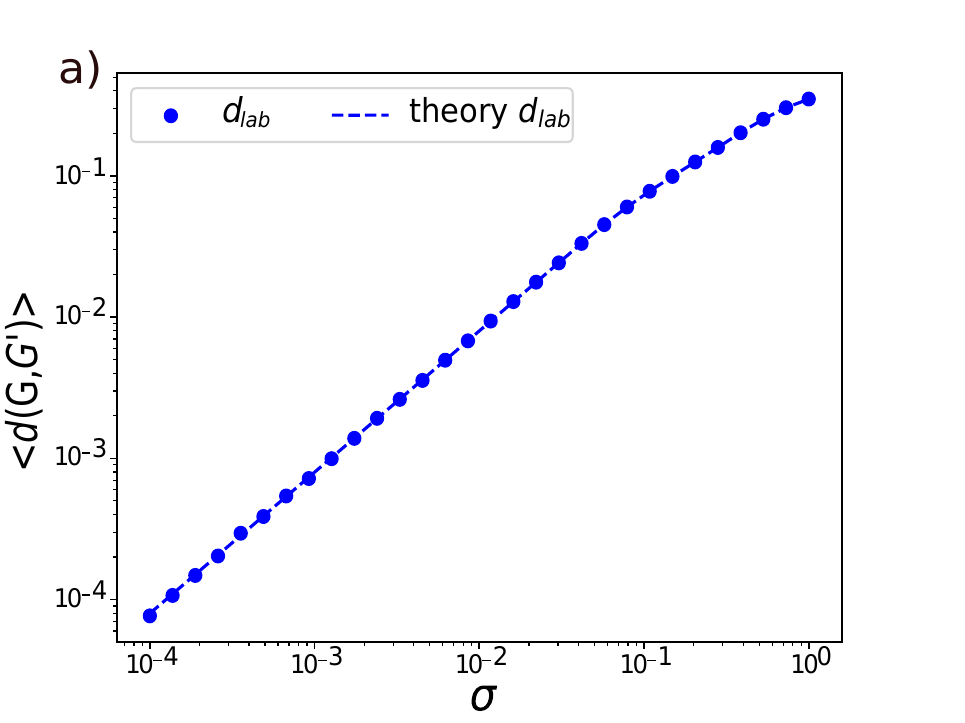}
\includegraphics[width=0.45\linewidth]{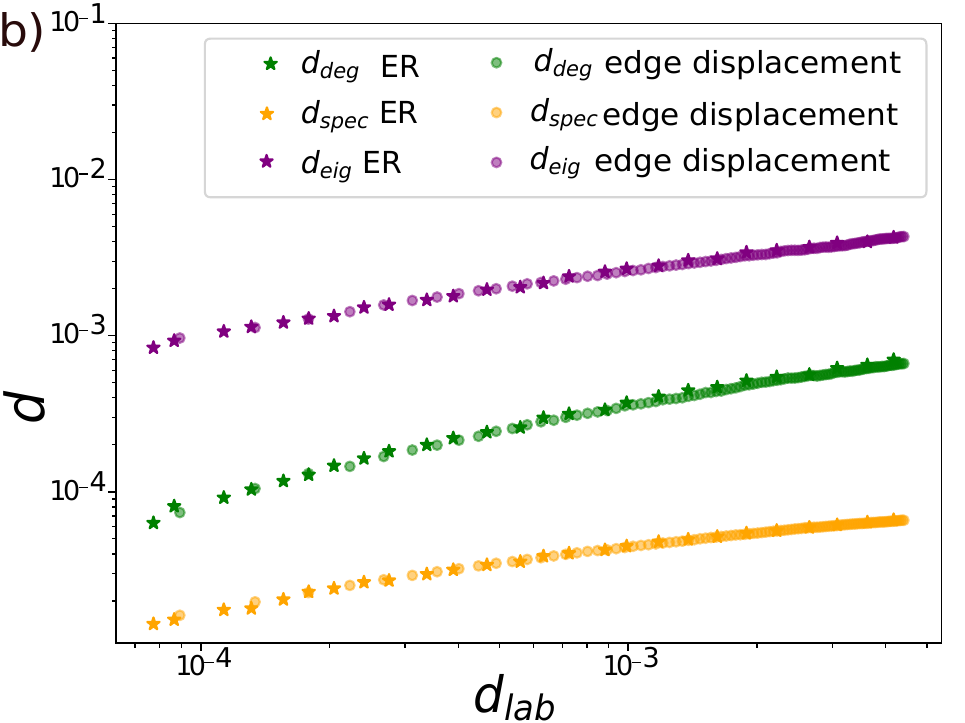}
\caption{\textbf{Permutation-invariant pseudo-distances and distance between labelled networks} 
 {Panel (a) shows numerical results for \(\langle d_{\text{lab}}(G,G') \rangle\) as a function of \(\sigma\) (solid circles), see the text for details. The theoretical prediction of Eq.~\eqref{Eq:d_ED_in_perturbations} is shown in dashed lines.  Panel (b) shows log-log plots of scatter plots between the three pseudo-distances $d_X$ and the labelled distance \(d_{\text{lab}}\), obtained numerically from the models described in Secs.~\ref{sec:er} and ~\ref{sec:displace}: data shown as stars is from perturbing Erd\H{o}s-R\'enyi graphs (Sec.~\ref{sec:er}), while solid dots are obtained in the model that displaces edges (Sec.~\ref{sec:displace}). The different colours represent the three different pseudo-distances. The networks consists of 300 nodes and an average of 5000 edges, and each data point is the average over an ensemble of 100 realisations. We observe that the scatter plots obtained via the two different procedures systematically collapse (for each given pseudo-distance). Within the tests performed here, the relation of any given pseudo-distance to the labelled distance appears to be continuous, monotonically increasing, and non-linear.}}
\label{fig:small_perturbation}
\end{figure}

\subsubsection{Perturbation using Erd\H{o}s-R\'enyi graphs}\label{sec:er} 

We start by constructing a reference graph $G$, which we will then later perturb to construct another graph $G'$ (in the notation of the previous section, $G'\equiv H$). To generate the reference graph, we draw a total of $N(N-1)/2$ standard uniform random variables $x_{ij} \in U(0,1)$, one for each pair of nodes $i<j$. We then fix a parameter $p$ to a value between zero and one. Nodes $i$, $j$ are linked in $G$ if and only if $x_{ij}<p$. Thus, $G$ is simply an Erd\H{o}s-R\'enyi graph with parameters $(N,p)$.\\
Subsequently, to build the graph $G'$ we `perturb' the numbers $x_{ij}$. More specifically, we introduce $x'_{ij} = x_{ij}+\xi_{ij}$, where the $\xi_{ij}$ are {\em iid} Gaussian random numbers with mean zero and variance $\sigma^2$. Nodes $i$ and $j$ are linked in $G'$ if and only if $x_{ij}'<p$. The quantity $\sigma$ is the model's control parameter, and intuitively it is easy to see that larger values of $\sigma$  make the perturbation on $G$ to be larger, and thus makes $G'$ `more different' than $G$.\\ 

We now quantify this using a proper distance between $G$ and $G'$. To do that, we initially assume that $G$ and $G'$ are labelled. We can then use Eq.~(\ref{eq:lab}) [in this example we set ${\cal Z}=N(N-1)$: this is the maximum value the sum on the right-hand side of that equation can take for graphs of size $N$].  We can compute the expected distance $d_{\text{lab}}(G,G')$ as a function of $\sigma$ analytically. This means to average over the ensemble of initial graphs $G$ (with fixed parameter $p$) and all perturbations $\xi_{ij}$. The resulting mean distance will depend on the parameter $p$ and on the size of the perturbation, $\sigma$. We find (see Appendix \ref{App:ED_small_perturbations}) that \begin{eqnarray}\label{Eq:d_ED_in_perturbations}
d_\text{lab}(\sigma)&=& 
\sigma \sqrt{\frac{2}{\pi}} \left[1 - \exp\left(-\frac{p^2}{2\sigma^2}\right) - \exp\left(-\frac{(p - 1)^2}{2\sigma^2}\right)\right] \nonumber \\
&& + \frac{1}{2} \left[1 - p \cdot \mathrm{erf}\left(\frac{p}{\sqrt{2} \sigma}\right) - (p - 1) \cdot \mathrm{erf}\left(\frac{p - 1}{\sqrt{2} \sigma}\right)\right].
\end{eqnarray}
For small $\sigma$ this becomes
\begin{equation}\label{eq:d_ED_approx}
d_\text{lab}(\sigma) \approx \sigma \sqrt{\frac{2}{\pi}} + \cdots,
\end{equation}
where the dots represent terms containing factors of the type $\exp\left[-p^2/(2\sigma^2)\right]$, and which hence tend to zero faster than linearly as $\sigma\to 0$.
{The validity of Eqs.~(\ref{Eq:d_ED_in_perturbations}) and (\ref{eq:d_ED_approx}) is indeed confirmed in simulations, as demonstrated in Fig.~\ref{fig:small_perturbation}(a). Simulation data are averages over initial graphs $G$ and perturbed graphs $G'$. We have thus shown that the above construction, combined with varying $\sigma,$ allows us to parametrically generate pairs of graphs of designated expected distances $d_\text{lab}$.} Given that \( d_{\text{lab}} \) increases linearly with $\sigma$ for small \( \sigma \), we can now explicitly use \( d_{\text{lab}} \) as the ground true quantifier of the perturbation that distinguishes $G$ from $G'$.

\medskip \noindent
We are now ready to test the relation between the pseudo-distances $d_X(G,G')$ (which can be measured for the unlabelled versions of $G$ and $G'$) and the ground true $d_{\text{lab}}(G,G')$. 
We have performed extensive numerical simulations and computed $\langle d_{\text{deg}}(G,G')\rangle$, $\langle d_{\text{spec}}(G,G')\rangle$ and $\langle d_{\text{eig}}(G,G')\rangle$ for different choices of $\sigma$ (the angle bracket denotes an average over realisations of the graphs $G$ and $G'$). In Fig.~\ref{fig:small_perturbation}(b) we scatter-plot $\langle d_X(G,G') \rangle$ against $\langle d_{\text{lab}}(\sigma)\rangle $ (the plot is on a log-log scale). We find a monotonically increasing relation, and no discernible discontinuities. However, this relation is not linear.

\subsubsection{Perturbation through link displacement}\label{sec:displace}
A method of constructing a set of graphs $\{H_0,\dots,H_L\}$ with designated pairwise distances (and with $L$ an integer parameter) was proposed in \cite{lacasa2022correlations, caligiuri2023}.
We begin by generating an Erd\H{o}s-R\'enyi network $H_0$ of size $N$ and with parameter $p$. To construct $H_1$ we displace one edge in $H_0$ to a previously vacant location (i.e., we make sure that no double edge is created during the displacement).  Subsequently, $H_2$ is constructed from $H_1$ by displacing a further edge. This edge is chosen different from the edge that was displaced in the construction of $H_1$ from $H_0$, and the place it is displaced to is chosen such that neither $H_0$ nor $H_1$ have a edge there. This process continues iteratively until $H_L$ has been constructed. In each step $H_n$ is constructed from $H_{n-1}$ by displacing an edge that is present in all graphs, $H_0,\dots,H_{n-1}$ to a location that is vacant in all graphs $H_0,\dots,H_{n-1}$. Further below in Sec.~\ref{subsubsec:Low} we will refer to the set $H_0,\dots, H_L$ as a `dictionary' of graphs.

\medskip \noindent
Using Eq.~(\ref{eq:lab}), it is easy to see that 
\begin{equation}
    d_\text{lab}(H_n,H_m)=\frac{2}{{\cal Z}}\vert n-m \vert
\end{equation} 

In the context of this example, we choose ${\cal Z}=2L$, so that the maximal distance between graphs in the set $H_0,\dots, H_L$ is $d_\text{lab}(H_0,H_L)=1$.

\medskip \noindent
The above protocol allows us to construct pairs of graphs of different distances $d_\text{lab}$. We can then measure the pseudo-distances $d_{\text{deg}}, d_{\text{spec}}$ and $d_{\text{eig}}$ between these pairs. Varying $\vert n - m\vert$ we can then again produce a parametric plot showing the pseudo-distances between pairs as a function of the distance $d_\text{lab}$. 

\medskip \noindent
Results are again shown in Fig.~\ref{fig:small_perturbation}(b). Each marker is an average over realizations of pairs of graphs with the same $d_\text{lab}$.
The data in the figure shows that the relation between $d_X$ and $d_\text{lab}$ is the same as that found in Sec.~\ref{sec:er}. This apparent universality is, however, likely due to the fact that the networks we have used in Sec.~\ref{sec:er} and in the current section have the same number of nodes, and a similar average number of edges. Further simulations (not shown here) reveal that the broad shape of the relationship between any one of the pseudo-distances and $d_\text{lab}$ remains similar when the network size or the edge density is varied, but that the specific details vary (i.e. the parameters of a possible fit).  Accordingly, while all functional relationships between $d_X$ and $d_\text{lab}$ appear to be continuous, non-decreasing and non-linear, we cannot establish a simple and universal quantitative mapping of the form $d_X = f_X(d_{\text{lab}})$.  

\medskip \noindent As a summary, numerical evidence suggests that the pseudo-distances $d_X$ defined above are well-behaved in the sense that they show a monotonic dependence with the ground true distance $d_{\text{lab}}$, and we have not observed discontinuities. 
As we will describe below, this opens up the prospect of generalizing network autocorrelation functions to unlabelled graphs. 
The lack of a linear dependence between $d_X$ and $d_{\text{lab}}$, however, suggests that an existing ground-truth sensitivity to initial conditions in a labelled temporal network (marked by exponentially growing labelled distance) cannot necessarily be detected as exponential expansion of $d_X$ from the unlabelled network trajectory. In other words, chaotic unlabelled networks do not necessarily display exponential divergence of nearby conditions, and a more careful analysis is required.

\section{Quantifying unlabelled network trajectories}\label{Sec:metrics}

\subsection{Sensitivity to initial conditions}\label{SubSec:MLE}

In this section, we will analyze how the pseudo-distances introduced in Sec.~\ref{Sec:Distances} behave in cases where the original (labelled) network dynamics is chaotic \cite{caligiuri2023}. Sec.~\ref{subsubsec:Low} deals with models of low-dimensional chaotic temporal networks, whereas the high-dimensional case is treated in Sec.~\ref{subsubsec:High}.

\subsubsection{Low-dimensional chaotic network dynamics}\label{subsubsec:Low}

The first analysis is centered on the dependence on initial conditions in an artificial temporal network that follows low-dimensional chaotic dynamics. Using the so-called {\it dictionary trick} (see \cite{lacasa2022correlations, caligiuri2023}), we map a one-dimensional chaotic time series --e.g. generated by the chaotic logistic map $x_{n+1}=4x_n(1-x_n)$-- onto a (labelled) temporal network.
This is done by initially performing an homogeneous partition the unit interval into $L$ subintervals, finding the subinterval $\left(\frac{k}{L}, \frac{k+1}{L}\right]$ where each $x_n$ of the 1D chaotic trajectory belongs to, and symbolising $x_n$ as the graph $H_{k}$ of the dictionary defined in Sec.~\ref{sec:displace}.
We note that the Lyapunov exponent associated with this logistic map is $\ln 2$. 

\begin{figure}[htb]
\includegraphics[width=.5\linewidth]{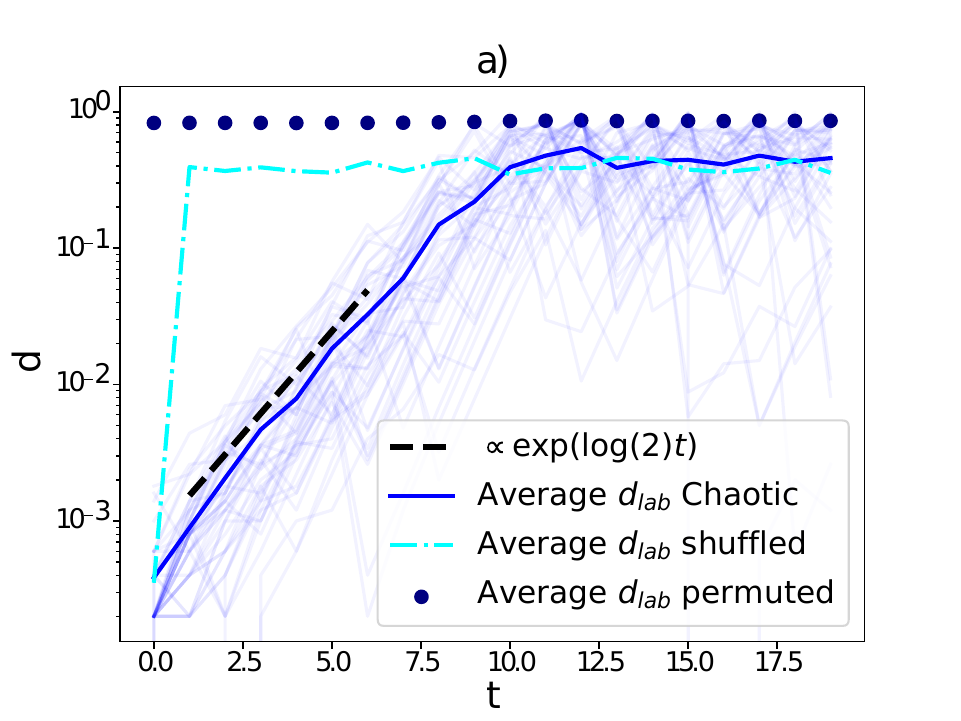}
\includegraphics[width=.5\linewidth]{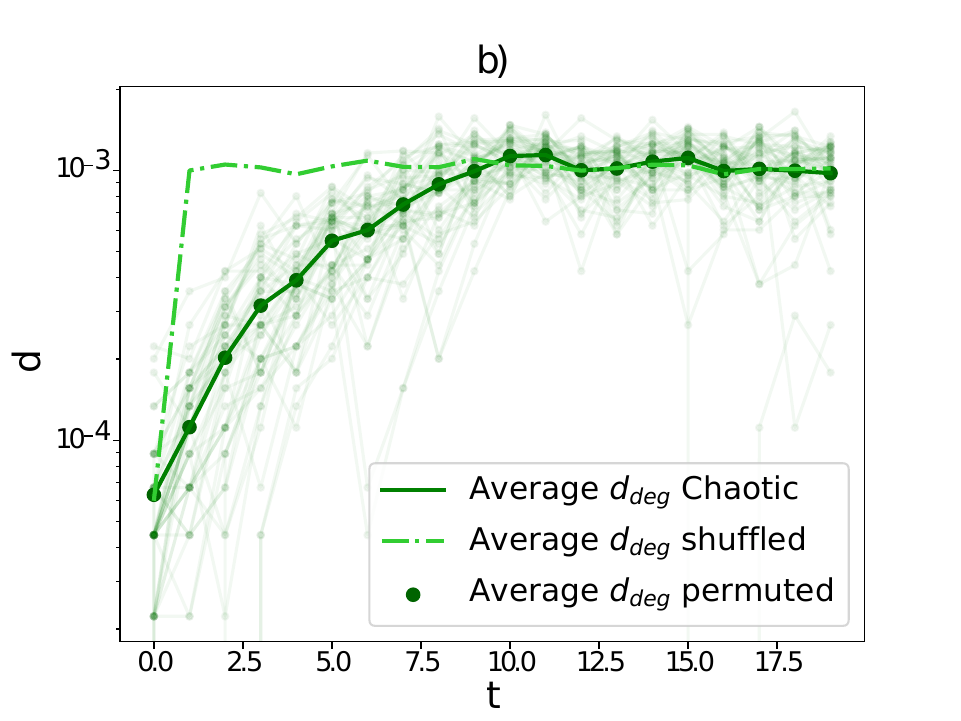}
\includegraphics[width=.5\linewidth]{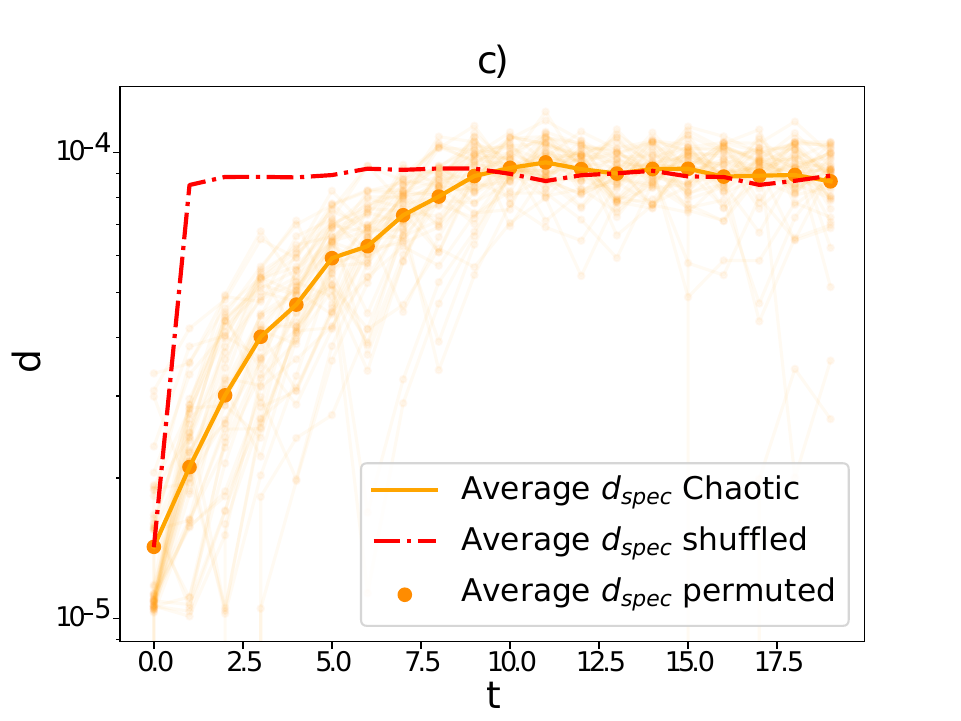}
\includegraphics[width=.5\linewidth]{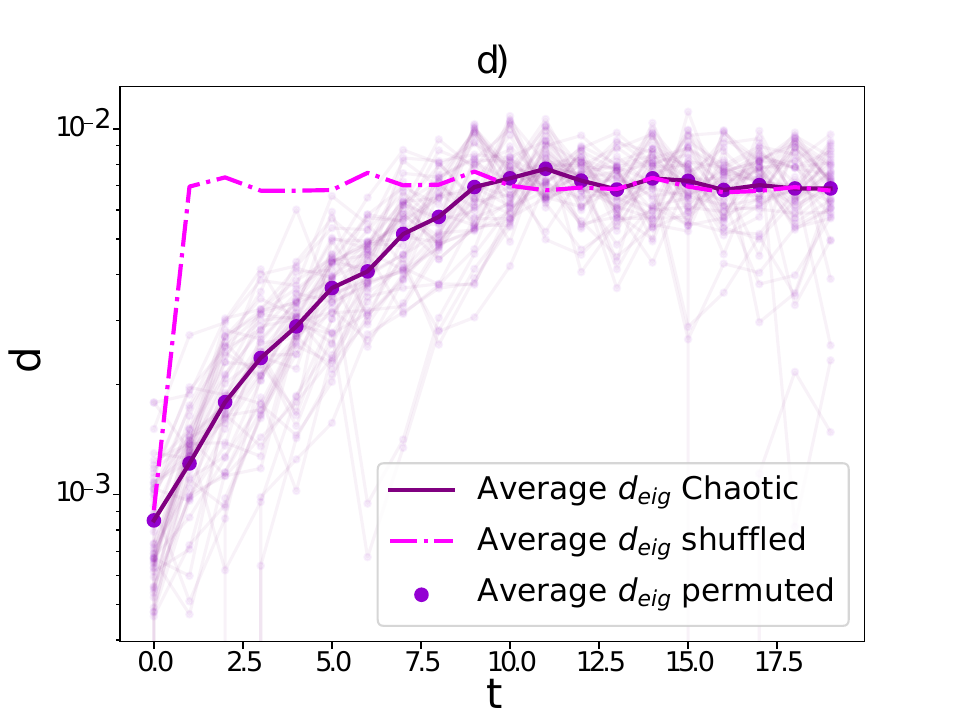}
\caption{{\textbf{Time-evolution of pseudo-distances of initially close (unlabelled) networks evolving via a low-dimensional chaotic dynamics.} Panels [a-d] depict semi-log plots of the different  distances of initially close trajectories over time. The network dynamics are generated via a fully chaotic logistic map using the dictionary trick, with a dictionary of $L=5\cdot 10^3$ networks, where each network has $N=300$ nodes (see the text in \ref{subsubsec:Low} for details). 
Each panel highlights in solid line the average distance (average over 50 pairs of close initial conditions), and in solid dots the equivalent result when node labels are shuffled, highlighting that only the pseudo-distances are invariant under random node labelling and are thus genuinely fit to address unlabelled network trajectories. Additionally, we also plot the result after shuffling the snapshot order (dash-dotted line).
 Panel (a), depicts (for comparison) the labelled distance  \( d_{\text{lab}}(t)\), that displays a clear exponential phase with an exponent close to $\ln 2$. Panels (b)-(d) depict the expansion as measured by 
the degree sequence pseudo-distance \( d_{\text{deg}}\), the spectral pseudo-distance \( d_{\text{spec}}\), and the eigenvector centrality pseudo-distance \( d_{\text{eig}}\) respectively. All pseudo-distances are capable of capturing the sensitivity to initial conditions although the growth is not strictly exponential, as expected.
}}
\label{fig:Low_chaos}
\end{figure}

\medskip \noindent
To measure sensitive dependence on initial conditions directly on the network trajectory, following \cite{caligiuri2023} we use a network version of Wolf's algorithm, identifying close recurrences in graph space and tracking the expansion of these over time, as shown in Fig.~\ref{fig:Low_chaos}(a). The labelled distance $d_\text{lab}(t)$, averaged over pairs of recurrent --i.e. close-- starting points, is shown as a solid blue line in that panel. We find exponential behaviour in an initial phase, before saturation is reached. This initial growth is of the form $d_{\text{lab}}\sim \exp(\lambda t)$ with an exponent $\lambda \approx \ln 2$, in agreement with the exponent of the logistic map.\\
For comparison, we also plot [in Fig.~\ref{fig:Low_chaos}(a)] the average labelled distance $d_{\text{lab}}(t)$ after a random permutation of node labels has been applied to each network snapshot (blue dots). This is to mimic the effect of forcing us to  use a labelled distance in an unlabelled trajectory, for which a node labelling is required. Random node labelling removes any temporal structure, and we thus find a flat curve {$d_{\text{lab}}(t)\approx 0.85$}. We have also shuffled the network trajectory (this means shuffling the order of the snapshots), and the resulting $d_{\text{lab}}(t)$ also shows a flat shape [cyan curve in Fig.~\ref{fig:Low_chaos}(a)], albeit with a smaller constant, probably due to the fact that shuffling snapshots is a less severe intervention than permuting node labels.

\medskip \noindent
Then, in Fig.~\ref{fig:Low_chaos} (b)-(d) we repeat the analysis but now using the three pseudo-distances $d_X$. By construction, these work directly with the unlabelled network trajectory and therefore do not use node labels. For all three pseudo-distances we find an expansion phase, i.e. the pseudo-distances qualitatively capture this fingerprint of chaotic behavior that $d_{\text{lab}}$ cannot retrieve from the unlabelled time series. However, the expansion of the unlabelled distances is not strictly exponential and therefore the agreement is not quantitative. This is actually expected, and in-line with the lack of linearity in the relation between the pseudo-distances and $d_{\text{lab}}$ observed in Sec.\ref{Sec:Distances}.

\medskip \noindent
In Sec.~\ref{Sec:Distances} we also found indications that it might be possible, for any {\it concrete} synthetic model of chaotic network trajectories, to find a monotonic yet non-linear function $f_X$ such that $d_X = f_X(d_{\text{lab}})$. If the labelled distance in the ground-truth labelled networks grows exponentially, $d_\text{lab}(t)=d(0)\exp(\lambda t)$, one then has $d_X(t)= f_X[d(0)\exp(\lambda t)]$, and therefore  $\lambda \approx (1/t)\ln f^{-1}_X [d_X(t)]$ (neglecting terms of the form $\mbox{const}\times t^{-1}$). However, if $f_X$ is model-dependent as it appears to be the case, we can only estimate this function on a case-by-case basis. Consequently, we currently cannot construct a well-defined (i.e. model-independent) Lyapunov exponent for unlabelled temporal networks which could be universally applicable.

\begin{figure}[htb]
\includegraphics[width=.5\linewidth]{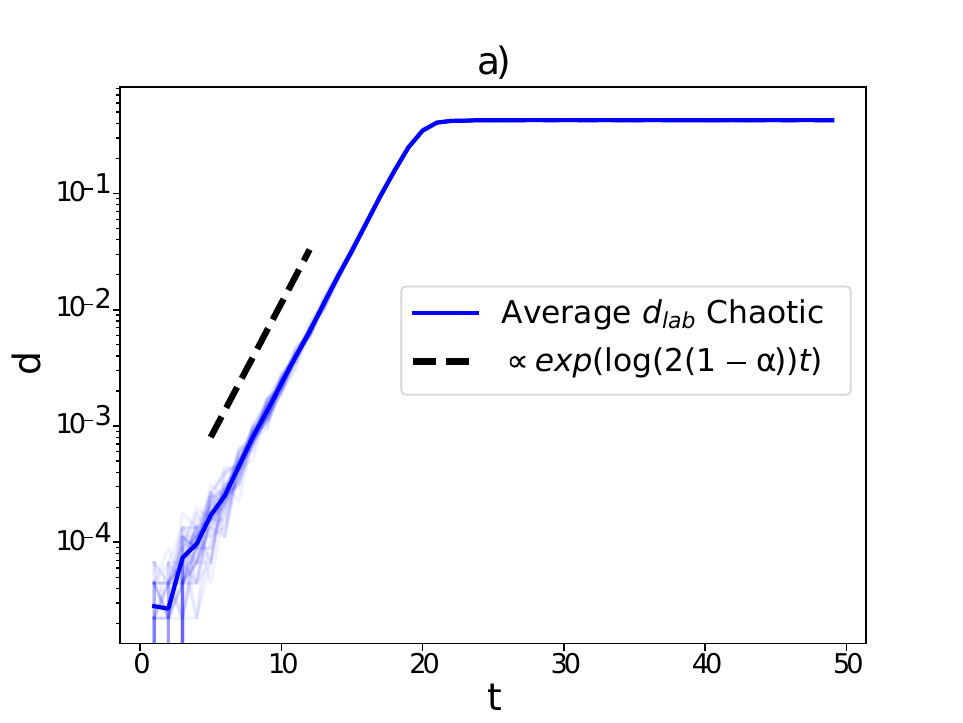}
\includegraphics[width=.5\linewidth]{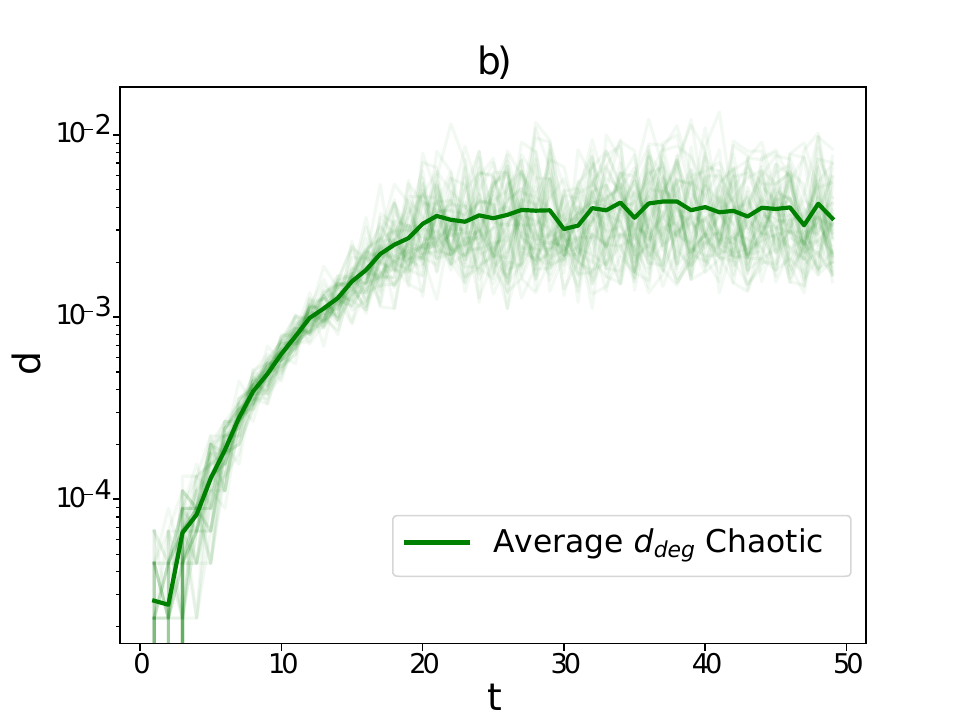}
\includegraphics[width=.5\linewidth]{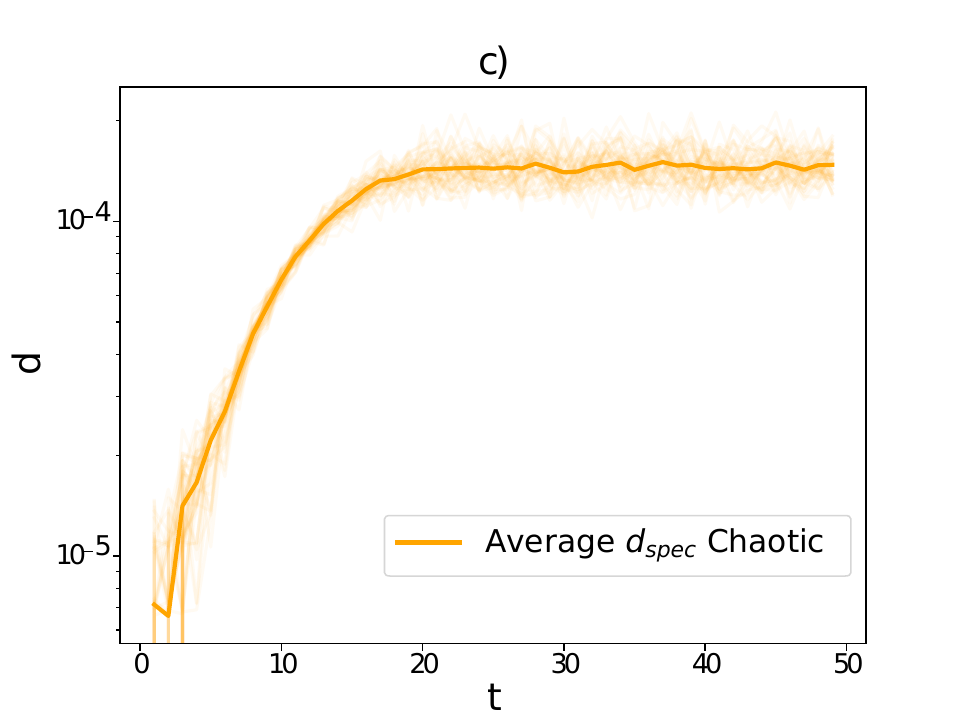}
\includegraphics[width=.5\linewidth]{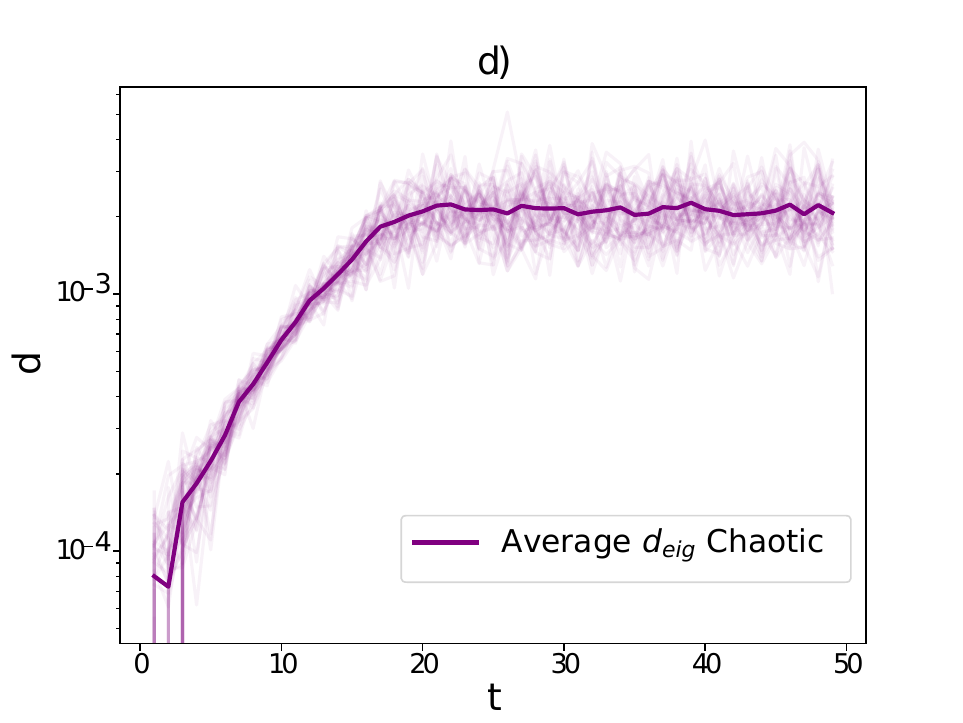}
\caption{\textbf{Time-evolution of pseudo-distances of initially close (unlabelled) networks evolving via a high-dimensional chaotic dynamics.} This figure is equivalent to Fig. \ref{fig:Low_chaos}, except here the dynamics is generated via a system of $m=N(N-1)/2$ globally coupled maps [Eq.~(\ref{eq:GCM}) with $\alpha=0.15$]. This system describes the (coupled) chaotic evolution of the edges of a network of $N=300$ nodes. Accordingly, the network trajectories display high-dimensional chaos (see Sec.~\ref{subsubsec:High} for details). When measured using the labelled trajectory, $d_\text{lab}(t)$ displays a clear exponential phase with an exponent close to the theoretical value $\ln 2(1-\alpha)$. When labels are lost, the pseudo-distances $d_X$ are still capable of capturing the expansion phase, although in these metrics such expansion is not strictly exponential.
}
\label{fig:High_chaos}
\end{figure}

\subsubsection{High-dimensional chaotic networks}\label{subsubsec:High}

To confirm the previous results on a higher-dimensional chaotic graph dynamics, we now repeat the same analysis for a network model constructed from a globally coupled map \cite{caligiuri2023} with a larger number $m=N(N-1)/2$ variables. Each of these variables corresponds to one (potential) edge $x_\ell$ in the network of size $N$, which we will construct. The coupled dynamics of the $\{x_\ell(t)\}_{\ell=1}^m$ variables is
\begin{equation}
x_\ell(t+1) = (1-\alpha)F[x_\ell(t)] + \frac{\alpha}{m}\sum_{\ell'=1}^m F[x_{\ell'}(t)], \ \ell=1,2,\dots,m, 
\label{eq:GCM}
\end{equation}
where $F(x)=1-2\vert x \vert$ is the chaotic tent map, and $\alpha\in [0,1]$ is a coupling constant. At each time $t$ we thus have $m$ real-valued variables $x_1(t), \dots, x_m(t)$, one for each potential edge in the network. {Once the time series for each variable have been simulated via Eq.~(\ref{eq:GCM}), we binarise these degrees of freedom such that edge $\ell$ exists at time $t$ if and only if $x_\ell(t) > \frac{1}{2}$}.
 This produces a network trajectory inheriting the chaotic properties of the globally coupled map \cite{caligiuri2023}. In particular this is a high-dimensional, turbulent dynamics in the weak coupling regime.

\medskip \noindent
In this model, since we have access to the dynamical equations from which the temporal network is generated, we can directly generate pairs of close initial conditions, and there is thus no need to look for recurrences in the temporal network. We then monitor the pseudo-distances $d_X(t)$ between pairs of  trajectories as they evolve in time. In Fig.~\ref{fig:High_chaos} (a) we show $d_\text{lab}(t)$, and in panels (b)-(d) the time evolution of the three pseudo-distances. Panel (a) confirms that the (labelled) network trajectory displays exponential expansion as measured by $d_\text{lab}(t)$, with a positive network Lyapunov exponent which coincides with the theoretical prediction for the globally coupled tent map \(\lambda = \ln [2(1-\alpha)] \) \cite{Kaneko,Morita1,Morita2}. Panels (b)-(d) of the same figure report the expansion as measured by the pseudo-distances $d_X$, when we deal with unlabelled trajectories which by construction cannot be monitored with $d_{\text{lab}}$.
 %As in the case of low-dimensional chaotic networks $d_{\text{lab}}(t)$ cannot capture the expansion phase once node labels have been randomised. This justifies using pseudo-distances [panels (b)-(d)]. 
 As in the low-dimensional case, while $d_X$ clearly display an expansion phase, the shape of this expansion is not strictly exponential.

\subsection{A Network Autocorrelation function of unlabelled network trajectories}\label{SubSec:Autocorrelation}
\subsubsection{Construction of metrics}
As a second way of characterizing the dynamics of unlabelled temporal network trajectories, we will construct a metric akin to an autocorrelation function. Autocorrelation functions can be used to detect periodicity in a time series or to quantify any dependence that the process has on its past (i.e., memory) \cite{Kantz_Schreiber_2003}.

In \cite{lacasa2022correlations}, a version of the autocorrelation function was proposed for labelled temporal networks. Both a matrix-valued autocorrelation function and a scalar autocorrelation function were introduced. Consider a labelled, unweighted temporal network with a fixed number of $N$ nodes given by a time-series of adjacency matrices $\{ \mathbf{A}(t) \}_{t=1}^T$. Here, $ \mathbf{A}(t) = \{ A_{ij}(t) \}_{i,j=1}^N $ is the $N\times N$ adjacency matrix of the $t$-th network snapshot. The scalar autocorrelation function $\tilde{{c}}(\tau)$ \cite{lacasa2022correlations} was defined as 
\begin{equation}\label{eq:ctilde}
\tilde{{c}}(\tau) := \texttt{tr} \bigg(\frac{1}{T - \tau} \sum_{t=1}^{T - \tau} \left[ \mathbf{A}(t) - \boldsymbol{\mu} \right] \cdot \left[ \mathbf{A}(t + \tau)^\top - \boldsymbol{\mu}^\top \right]\bigg ).
\end{equation}
In this expression $\mathbf{A}^\top $ denotes the transpose of the matrix $ \mathbf{A} $, and $\boldsymbol{\mu} = \frac{1}{T} \sum_{t=1}^T \mathbf{A}(t)$ is the annealed (i.e., time-averaged) adjacency matrix of the temporal network, and $\texttt{tr}(\cdot)$ is the trace operator. We will use $\tilde{c}(\tau)$ as the ground truth against which to compare our results once node labels are removed. We note that $\tilde c(\tau)$ is the sum of element-wise autocorrelation functions in the adjacency matrix as it can be written in the form $\tilde c(\tau)=\sum_{ij} \tilde c_{ij}(\tau)$, where $\tilde c_{ij}(\tau)=(T-\tau)^{-1}\sum_{t=1}^{T-\tau}\{ [a_{ij}(t)-\mu_{ij}][ a_{ij}(t+\tau)-\mu_{ij}]\}$.

\medskip \noindent 
For given $\tau$ the autocorrelation function of a signal can be interpreted as the scalar product of the signal --viewed as a vector of $T$ dimensions-- and itself, shifted by $\tau$. This scalar product is indicated by the sum over $t$ in Eq.~(\ref{eq:ctilde}). The autocorrelation function is consequently a similarity measure of the signal and the time-lagged signal.
With this in mind, we can then construct analogous similarity measures for unlabelled temporal networks, starting from the pseudo-distances $d_X$ defined in Sec.~\ref{Sec:Distances}. Consider a time series of unlabelled graphs, $G_1, G_2, \dots$. For any of the pseudo-distances $d_X$ ($X\in\{\mbox{deg, spec, eig}\}$) we then define
\begin{equation}
\label{eq:autocorrelation}
   c_X(\tau) := 1 - \frac{1}{T-\tau}\sum_{t=1}^{T-\tau} \frac{d_X(G_t, G_{t+\tau})}{{\cal J}}, 
\end{equation}
where 
\begin{equation}
{\cal J}= \frac{1}{T(T-1)}\sum_{t,t'=1}^T d_X(G_t, G_{t'})
\end{equation}
is an estimate of the average pseudo-distance between any pair of network snapshots in the trajectory.\\
For $\tau=0$ we have $d_X(G_t, G_{t+\tau})=0$ and thus $c_X(\tau=0)=1$, in-line with the properties of a conventional autocorrelation function. In the absence of memory in the time series and considering the limit of very large time lags $\tau$, it is also reasonable to assume that $\frac{1}{T-\tau}\sum_{t=1}^{T-\tau}d_X(G_t, G_{t+\tau})$ approaches the average distance $\cal J$ between any two graphs in the time series. We therefore expect $c_X(\tau)$ to approach the value zero for large $\tau$, similar to the behaviour of an autocorrelation function in systems without memory.\\
In the next subsections, we will explore the behavior of $c_X(\tau)$ on noisy periodic and non-periodic synthetic network trajectories with memory.

\subsubsection{Test on noisy periodic network trajectories}
We evaluate the autocorrelation-like functions $c_X(\tau)$
for the noisy periodic temporal network model previously introduced in \cite{lacasa2022correlations}. In this model, initially we construct a strictly periodic sequence of graphs. The period will be denoted as $T_\text{period}$.  We  sample $T_\text{period}$ independent Erd\H{o}s-R\'enyi graphs, each with $N$ nodes and with fixed parameter $p$. These graphs are denoted as $G_1, \dots, G_{T_\text{period}}$. This sets the (noiseless) time series during one period. The full periodic time series is then obtained by repeating (concatenating) this sequence over and over again. Mathematically, this means to define $G_{nT_\text{period}+s}=G_s$ for all $n=1,2,\dots$ and $s=1, 2, \dots, T_{\text{period}}$. 
In a second step we now `pollute' this periodic network trajectory with noise. For each snapshot $G_t$, with $t>T_{\text{period}}$, each pair of nodes $i<j$ is independently selected for potential perturbation with probability $q$. Each edge sampled in this process is then set to be present in the perturbed graph with probability $p$ and absent with probability $1-p$. This occurs irrespective of whether the edge was present or not in $G_t$ before the perturbation.

\begin{figure}[htb]
\includegraphics[width=.5\linewidth]{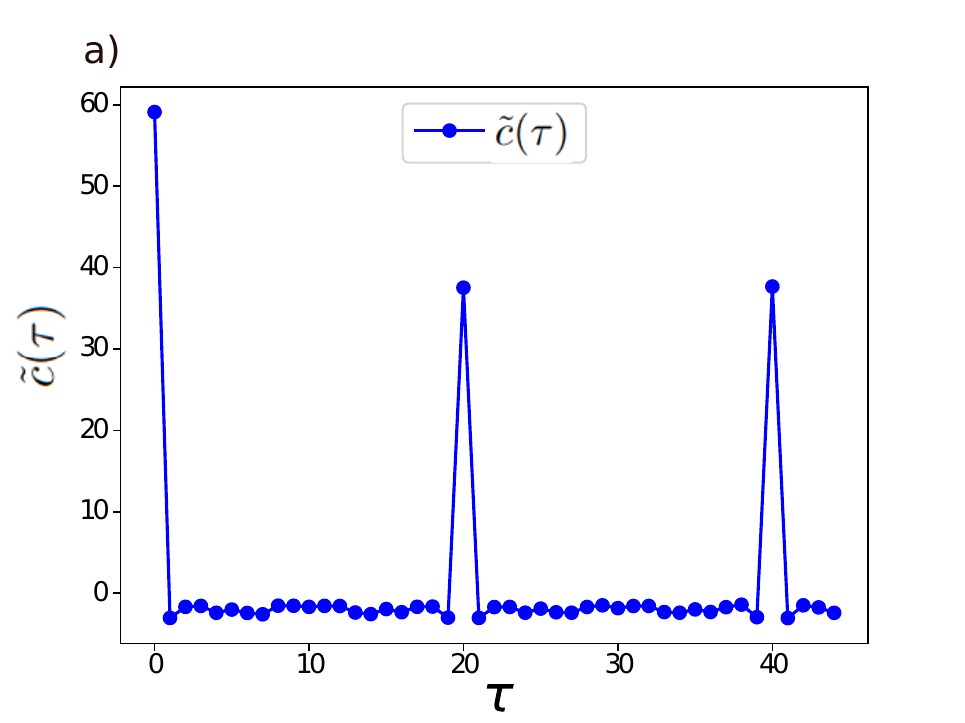}
\includegraphics[width=.5\linewidth]{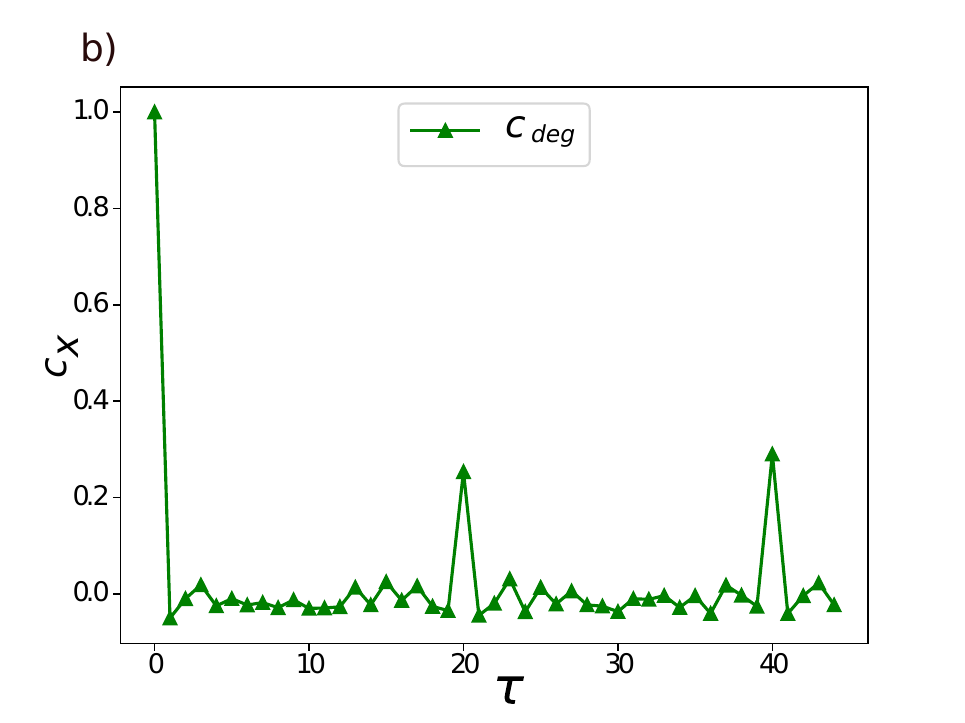}
\includegraphics[width=.5\linewidth]{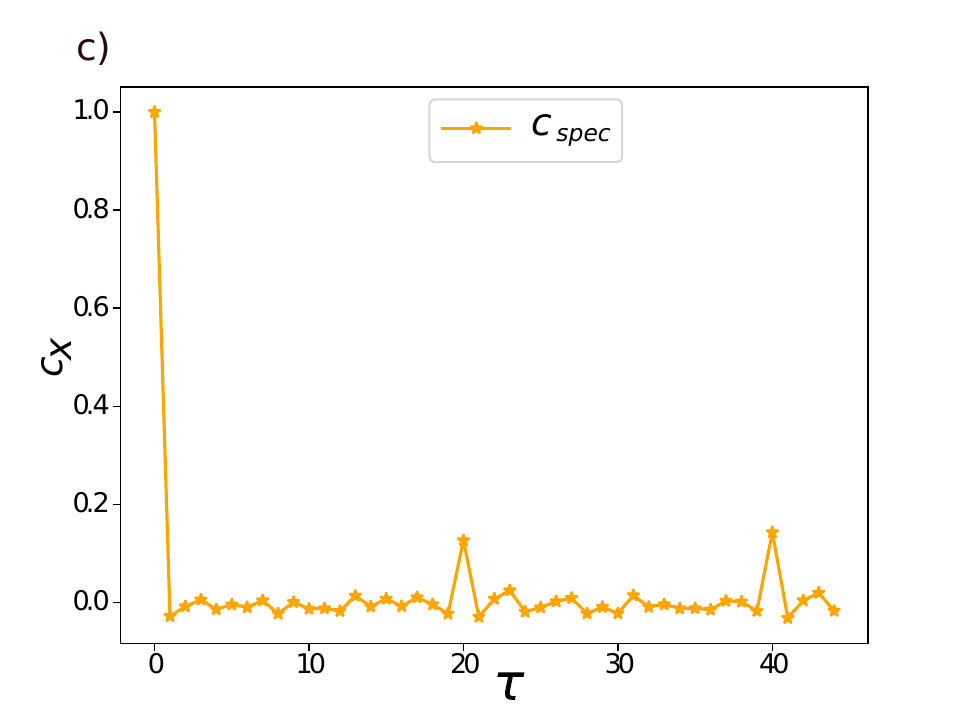}
\includegraphics[width=.5\linewidth]{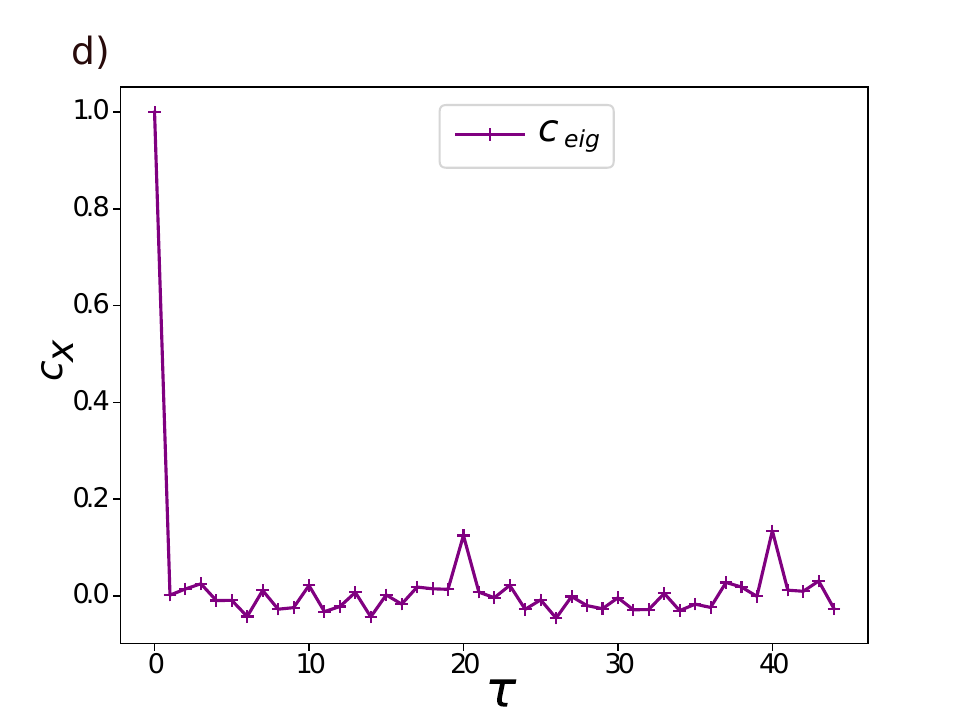}
\caption{\textbf{Autocorrelation functions for noisy periodic networks.} (a) Scalar autocorrelation function $\tilde{c}(\tau)$ (Eq.\ref{eq:ctilde}) applied to a (labelled) noisy periodic network trajectory of $T=400$ snapshots with parameters $N=20$, $T_{period}=20$, $q=0.2$, $p=0.2$.
In panels (b), (c), and (d) we plot the autocorrelation-like functions $c_\text{deg}(\tau)$, $c_\text{spec}(\tau)$, and $c_\text{eig}(\tau)$, respectively [Eq.~(\ref{eq:autocorrelation})] for the unlabelled version of the same network trajectory, where by construction $\tilde{c}(\tau)$ cannot be used. 
While the peaks indicating periodicity in panels (b), (c), and (d)  are lower compared to those in panel (a), the pseudo-distances can still detect periodicity in the unlabelled setting.}
\label{fig:Noisy_periodic}
\end{figure}

\medskip \noindent
In Fig. \ref{fig:Noisy_periodic}, we report results for the autocorrelation-like measures $c_X(\tau)$ 
for a noisy periodic network with $T_\text{period}=20$. For comparison, panel (a) depicts the scalar autocorrelation $\tilde{c}(\tau)$ obtained from the labelled temporal network. The peak at $\tau=20$ can be clearly seen. Results from the pseudo-distances are shown in panels (b)-(d). The data confirms that all three pseudo-distances are capable of retrieving the true period. The measure based on the degree sequence seems to exhibit the highest peak [panel (b)].

\medskip
\noindent To further assess the ability of the $c_X(\cdot)$ to detect periodicity in increasingly noisy temporal networks, we define an effective {z-score} 
\begin{equation}\label{eq:zscore}
z_X = \frac{ c_X(T_\text{period})- \mu_X}{\sigma_X},
\end{equation}
where $\mu_X$ and $\sigma_X$ represent the mean and standard deviation of the set $\{c_X(1), c_X(2), \dots, c_X(T_\text{period}-1)\}$. Thus, for each choice of $X\in\{\mbox{deg,\,spec,\,eig}\}$ the quantity $z_X$ reports, in units of standard deviation, the difference between the height of $c_X$ at $T_\text{period}$ and the average up to $T_\text{period}$.
Accordingly, a higher $z$-score indicates better performance of the autocorrelation-like measure in detecting periodicity.

\begin{figure}[htb]
\includegraphics[width=.5\linewidth]{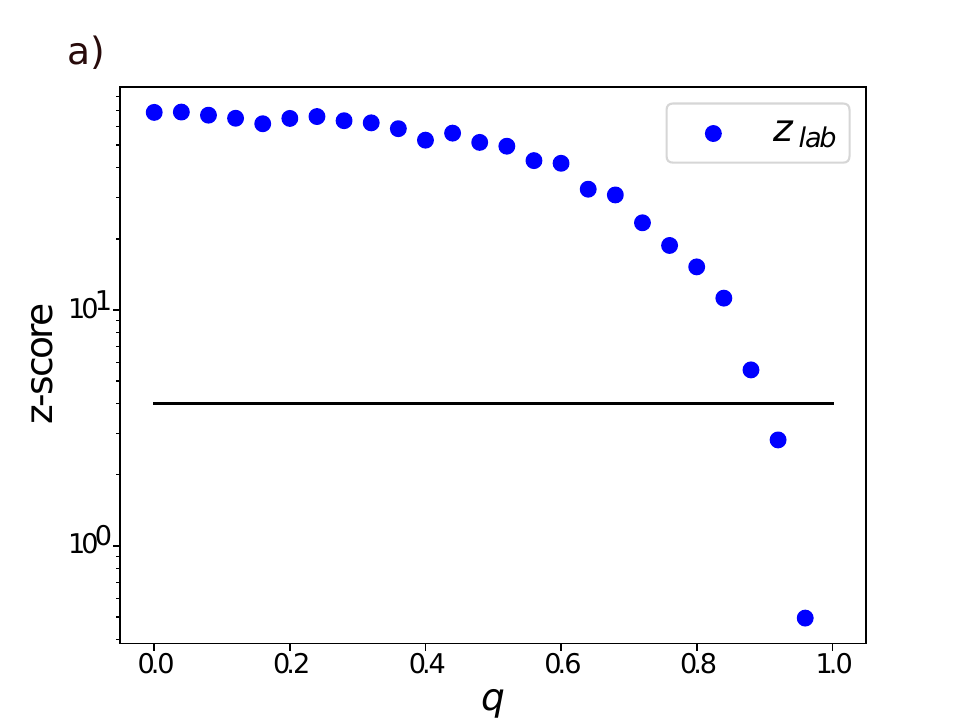}
\includegraphics[width=.5\linewidth]{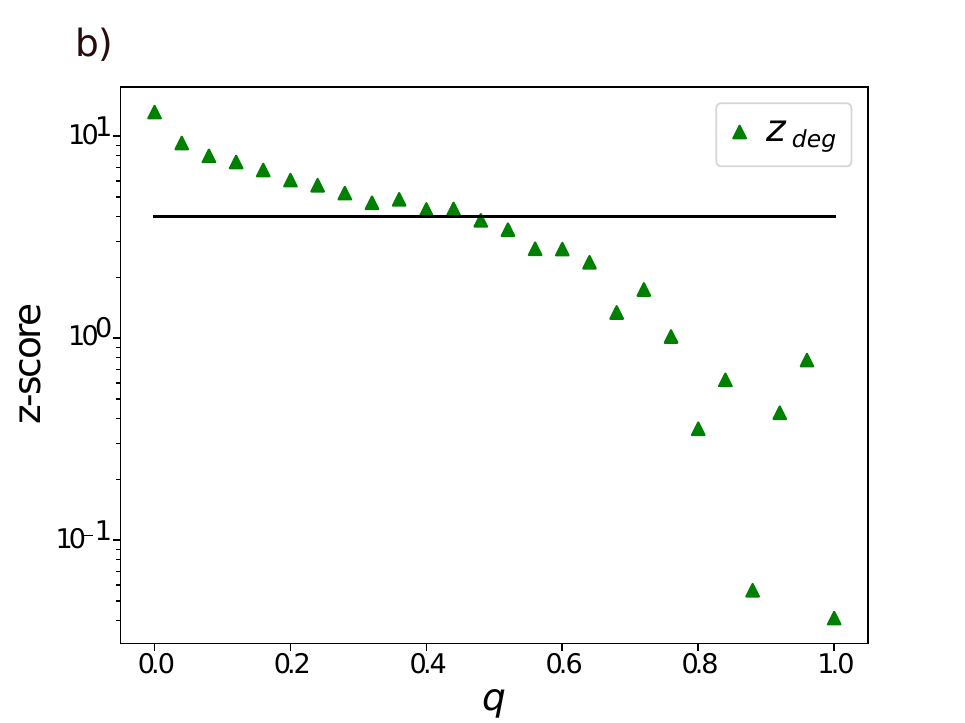}
\includegraphics[width=.5\linewidth]{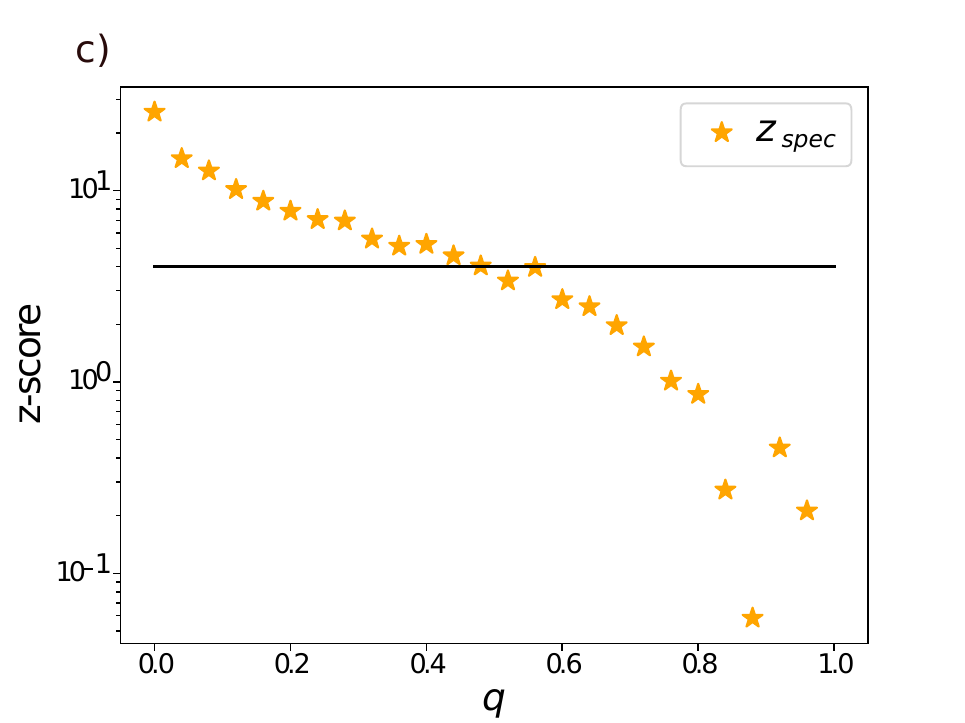}
\includegraphics[width=.5\linewidth]{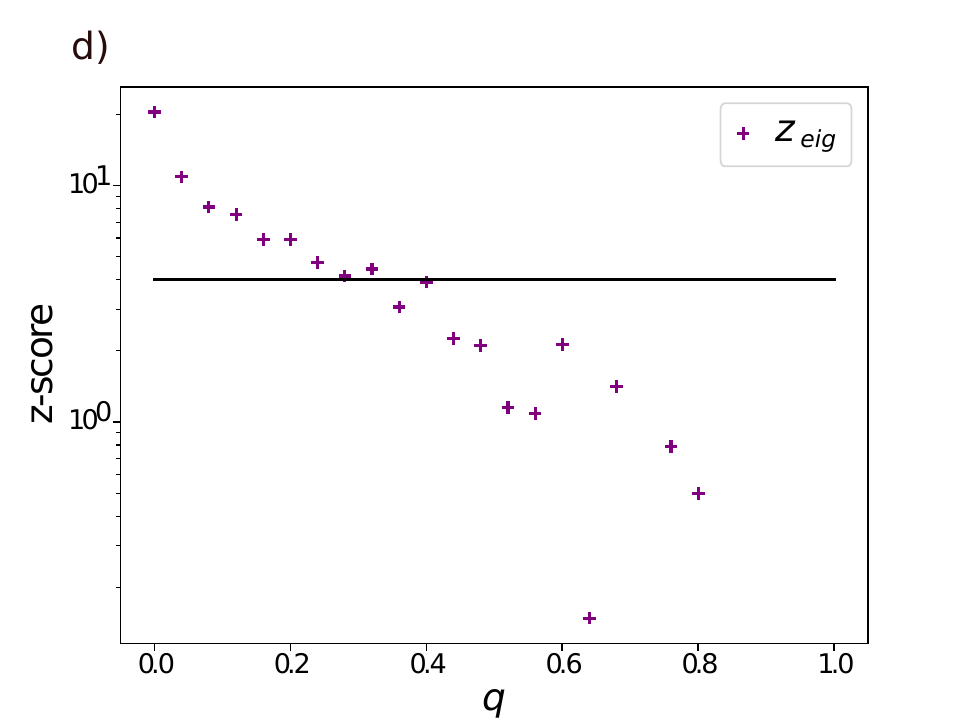}
\caption{{\textbf{Robustness against noise of the autocorrelation functions.} To assess the robustness of the autocorrelation functions $\tilde{c}(\tau)$ and the ones applicable to the unlabelled setting $c_X(\tau)$, we show the z-scores $z_X$ [Eq. \ref{eq:zscore}] as a function of the noise parameter $q$, as applied to noisy periodic network trajectories of $T=400$ snapshots with parameters  $N = 20$, $T_{\text{period}} = 10$ and $p = 0.2$ (data is averaged over $10$ realizations). Using the threshold $z>4$ as the criterion for period detectability, we see that all three $c_X(\tau)$ functions --universally applicable in labelled and unlabelled trajectories-- are robust against moderate amounts of noise [panels (b)-(d)], although comparatively less robust than $\tilde{c}(\tau)$ [panel (a)].}}
\label{fig:z_score}
\end{figure}

\medskip \noindent 
In Fig.~\ref{fig:z_score} we show $z_X$ as a function of the noise intensity $q$. Panel (a) shows results the autocorrelation function of the original labelled temporal network, and panels (b)-(d) are for the three different pseudo-distances. For easier interpretation, we also show a horizontal line at $z=4$ in each panel (indicating a peak four standard deviations higher than the rest of the autocorrelation function). We use this threshold as a conservative criterion to say that the method is able to detect periodicity in the network signal.

The ability of the unlabelled measures to detect noisy periodicity is seen to be weaker than that of the autocorrelation function for the labelled time series. The curves in panels (b)-(d) broadly cross the threshold level at $q\approx 0.4$, compared to $q\approx 0.9$ for the labelled case. Nonetheless, we note the ability of the unlabelled measures to detect periodicity at low and moderate noise levels. We also note that the metric based on eigenvector centrality $c_{\text{eig}}(\cdot)$ [panel (d)] is less robust to noise than the ones based on the degree sequence and the spectrum of the adjacency matrix.

\subsubsection{Test on Discrete Autoregressive Unlabelled network trajectories}
We now test the $c_X(\tau)$ on a model that generates network trajectories with memory. Specifically, we use a `Discrete Autorregressive Network model' with memory order $\rho$, or DARN($\rho$) \cite{Williams_2019}. We consider networks with $N$ nodes. In this model, the presence or absence of any given edge $\ell=1,\dots, N(N-1)/2$ in the graph at time $t$ is governed by a binary random variable $x_{\ell,t}\in\{0,1\}$. The random variables describing different edges are independent from one another, and are constructed as follows:
 At time $t$, each $x_{\ell,t}$ takes its value independently of the past history (this happens with probability $1-q$), or samples a state uniformly from one of its past $\rho$ states (this occurs with probability $q$). This can be written as follows,
\begin{equation}
x_{\ell,t} = Q_{\ell,t} x_{\ell, t-Z_{\ell,t}} + (1-Q_{\ell,t})Y_{\ell,t},
\end{equation}
where $Q_{\ell,t}$ is a Bernoulli random variable taking value one with probability $q$, and value zero with probability $1-q$. The quantity $Y_{\ell,t}$ is also a Bernoulli variable, with parameter $p$. Finally, $Z_{\ell,t}$ is a discrete random variable taking values in interval $\{1, \ldots, \rho\}$ with uniform distribution. 

\begin{figure}[htb!]
\includegraphics[width=.5\linewidth]{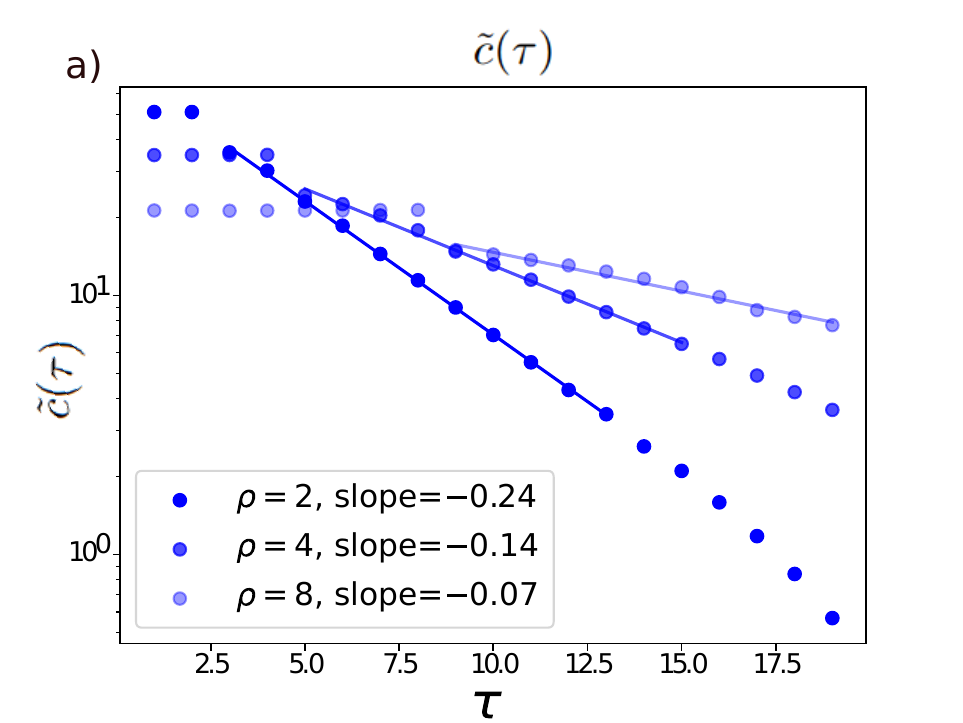}
\includegraphics[width=.5\linewidth]{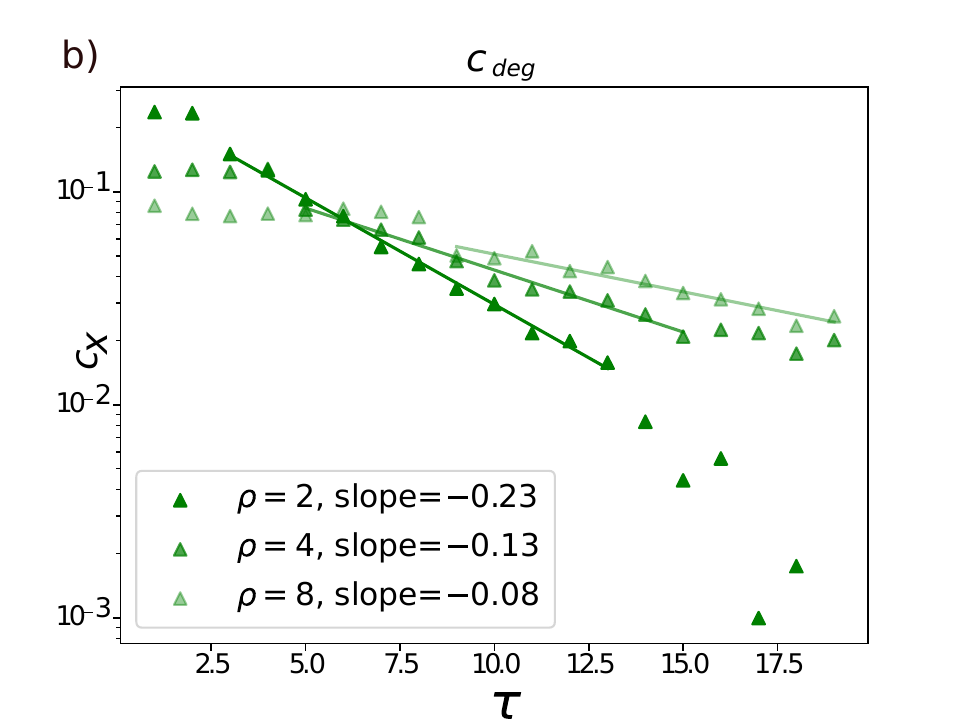}
\includegraphics[width=.5\linewidth]{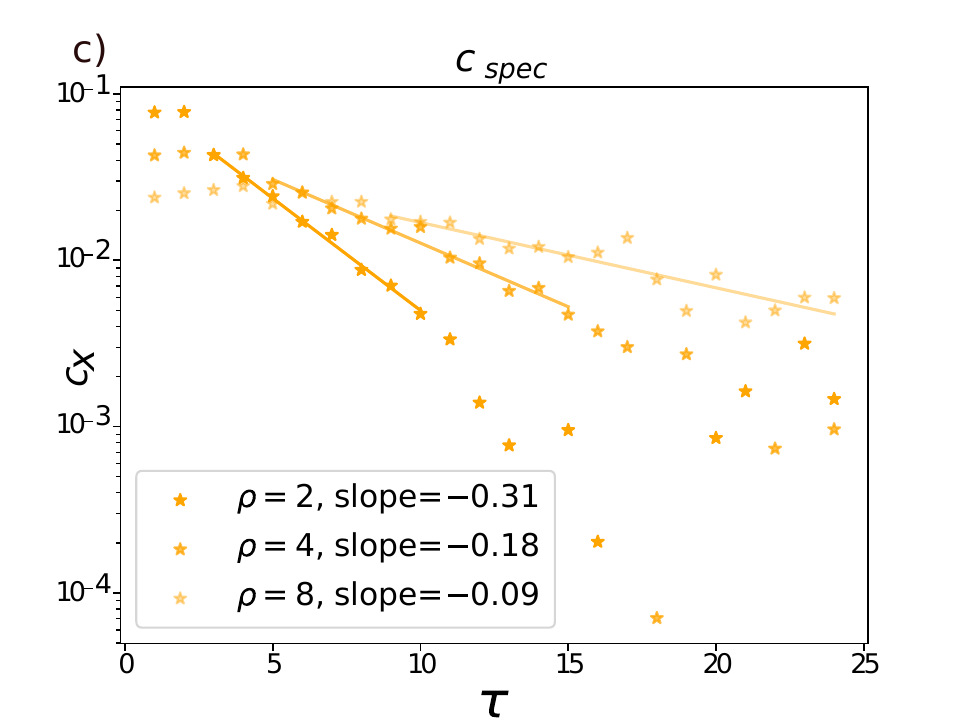}
\includegraphics[width=.5\linewidth]{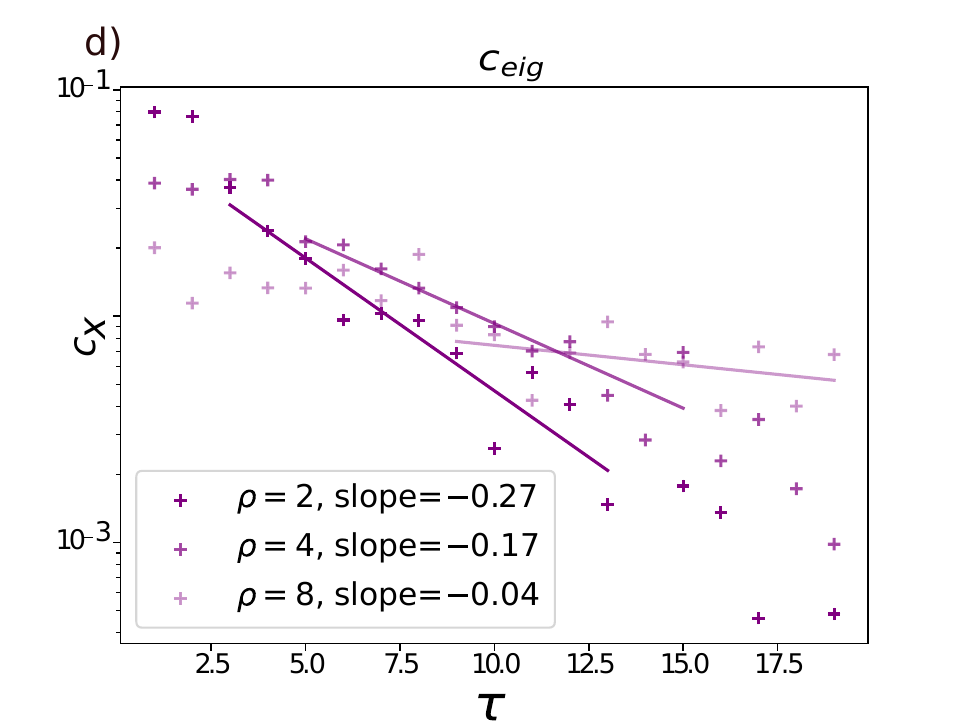}
\caption{\textbf{Autocorrelation-like function capture memory in DARN processes.} Semi-log plots of the scalar autocorrelation function $\tilde{c}(\tau)$ [panel (a)] and autocorrelation-like functions $c_{\text{deg}}(\tau)$, $c_{\text{spec}}(\tau)$, and $c_{\text{eig}}(\tau)$ [panels (b)-(d)] computed from a network trajectory of $T=10^3$ snapshots generated by a DARN($\rho$) process with parameters  $p=0.5$, $q=0.7$ and for $\rho=2, \ 4$ and $8$ (data is averaged over $10$ realizations). Solid lines are exponential fits. The typical flat shape for $\tau<\rho$, followed by an exponentially decaying curve is approximately preserved in $c_{\text{deg}}(\tau)$ and $c_{\text{spec}}(\tau)$ (the slope of the exponential decay is similar to the one in $\tilde{c}(\tau)$). Results for $c_{\text{eig}}(\tau)$ are more noisy.}
\label{fig:DARN}
\end{figure}

\medskip \noindent
In Fig.~\ref{fig:DARN}, we compare $\tilde{c}(\tau)$ and $c_X(\tau)$, for different temporal networks generated using the DARN process, varying the memory parameter $\rho$. As shown in panel (a), the  autocorrelation of the labelled DARN trajectories remains constant for $\tau \leq \rho$, and then decrease exponentially for $\tau > \rho$. The autocorrelation-like functions $c_X(\tau)$ for the unlabelled temporal networks [panels (b)-(d)] show similar behavior: they are approximately constant for $\tau \leq \rho$, and decay for $\tau > \rho$. The slope of the decrease of $c_\text{deg}$ is very similar to that of the labelled case [compare panels (a) and (b)]. The data for the metric based on $d_{\text{eig}}$ [panel (d)] is noisy, consistent with our previous observations in Fig.~\ref{fig:z_score}. We conclude that the measure based on this pseudo-distance has a comparatively lower capacity of detecting temporal correlations.

\subsection{Empirical (unlabelled) temporal network trajectories}\label{Sec:RealUN}

Here we finally test the autocorrelation-like functions $c_X(\tau)$ on three real-world temporal networks: US domestic flights  \cite{williams2016spatio}, face-to-face contact in a Malawi Village \cite{ozella2021using}, and spatial proximity of participants the 2009 Annual French Conference on Nosocomial Infections (SF2H, also referred to as SFHH) \cite{Genois2018}. In the latter dataset each node is uniquely identified by an anonymous ID. Links are  established whenever two individuals are detected in close proximity by RFID sensors. These interactions are considered active during 20-second intervals, i.e., the sensor receives signals every 20 seconds, making the network temporal. The same technology was also used to detect face-to-face contact in a village in Malawi with 86 individuals. The data on flights is from passenger flights within the USA during February 2014. Each node represents a U.S. airport, and a link is created when a flight connects two airports. The link remains active for the entire duration of the flight. For these three examples we have access to the node labels. We deliberately choose these temporal networks so we can verify if an analysis of the corresponding unlabelled temporal networks (i.e., after removal of labels) still recovers some of the features seen in the labelled time series. In this sense, this section is part of a validation procedure for our quantitative measures for unlabelled temporal networks.

\begin{figure}[htb!]
\includegraphics[width=.5\linewidth]{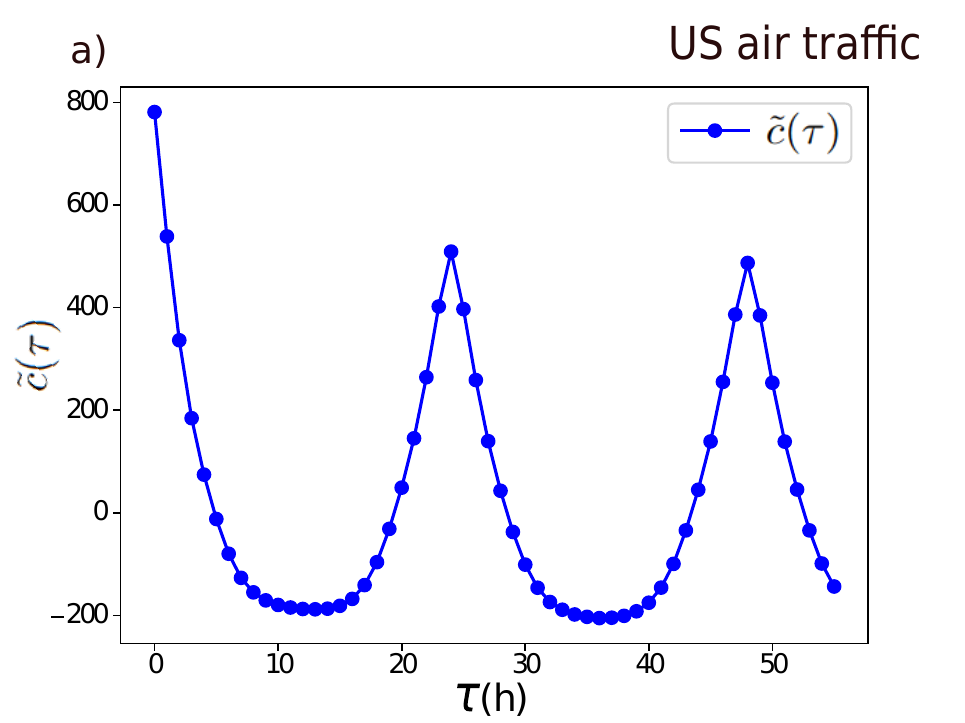}
\includegraphics[width=.5\linewidth]{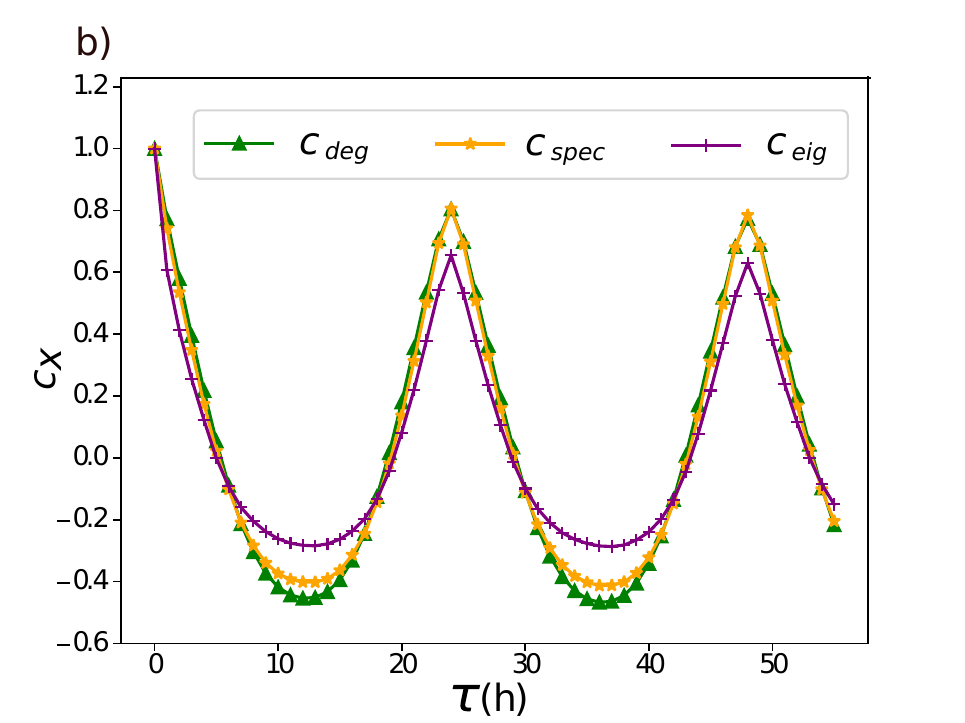}
\includegraphics[width=.5\linewidth]{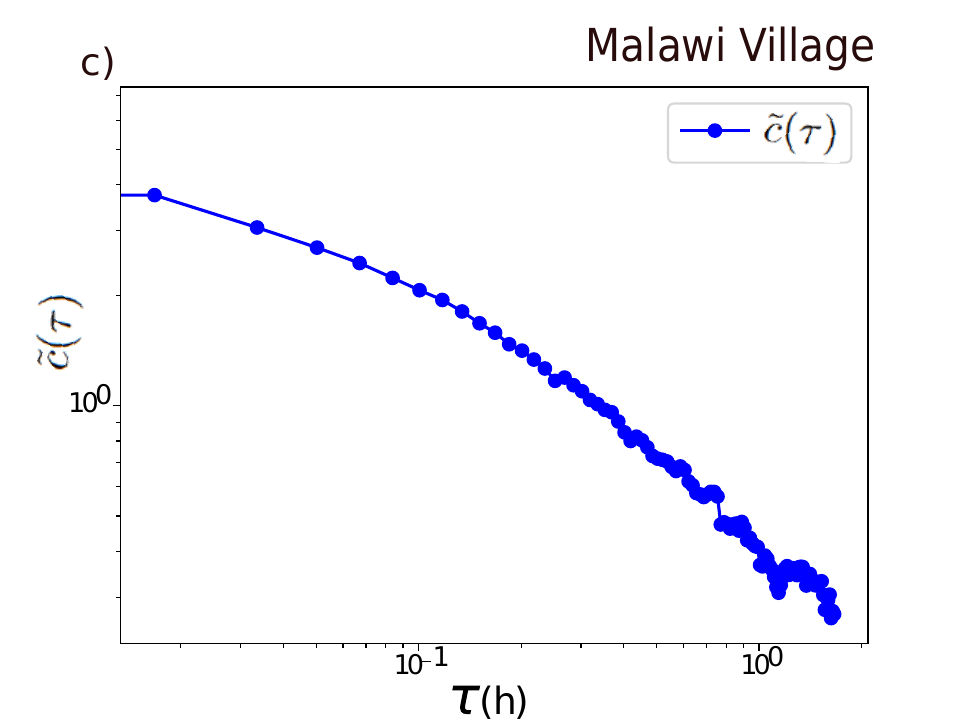}
\includegraphics[width=.5\linewidth]{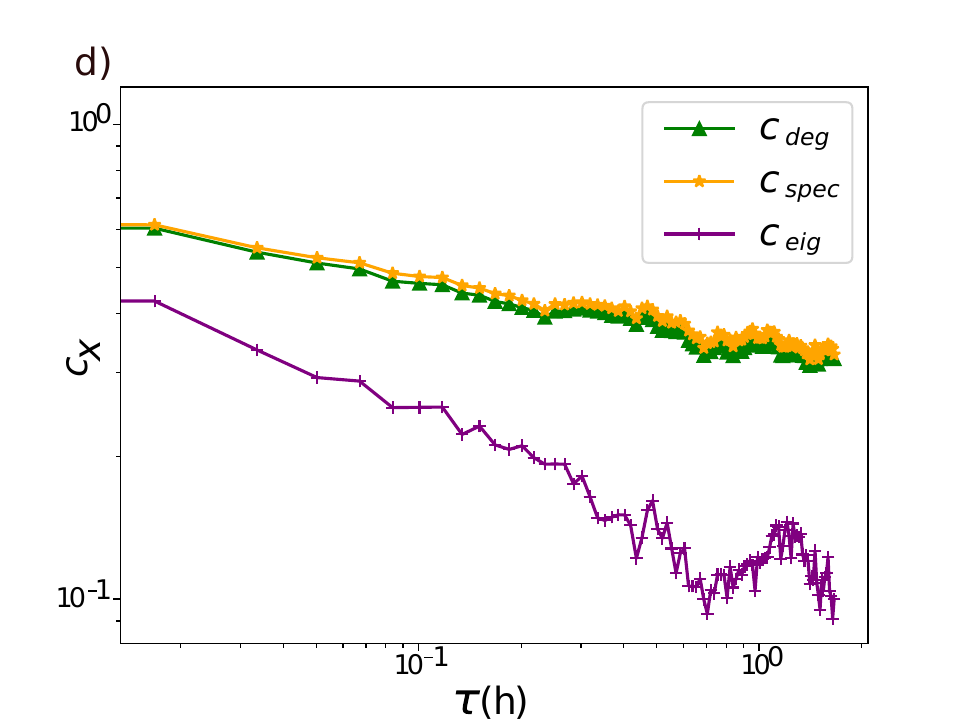}
\includegraphics[width=.5\linewidth]{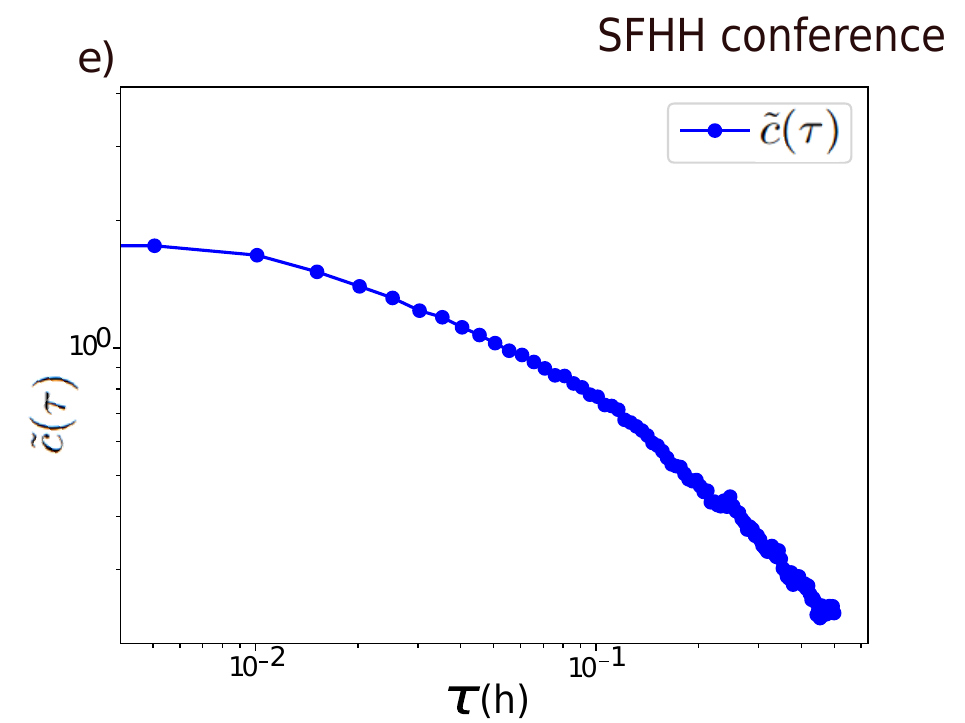}
\includegraphics[width=.5\linewidth]{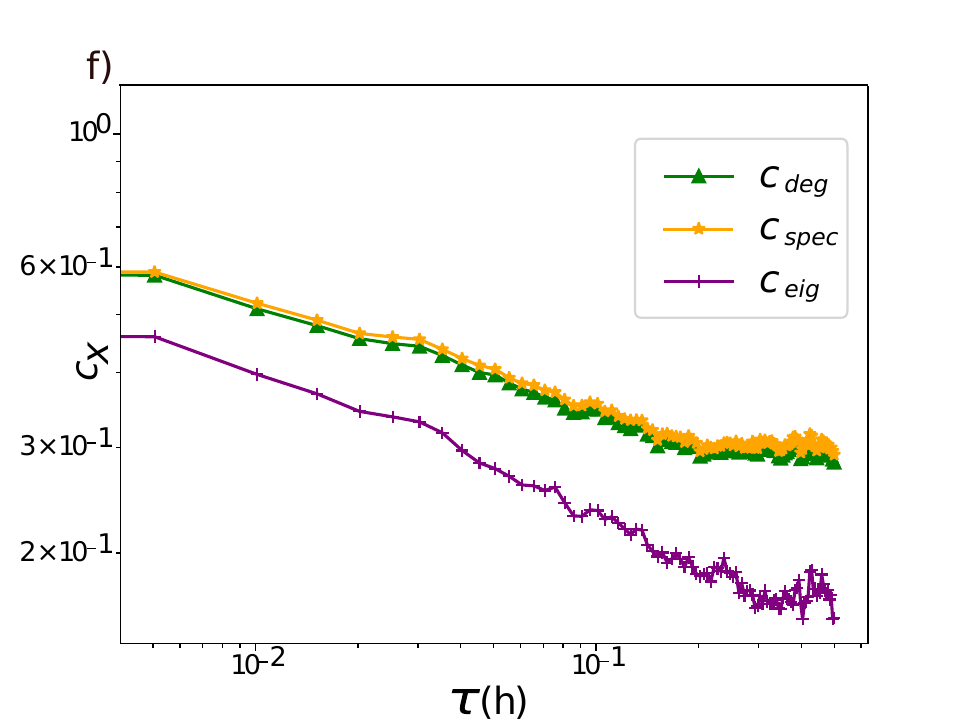}
\caption{{\textbf{Autocorrelation-like function for real-world empirical temporal networks.}  In panels (a), (c), and (e), we present log-log plots of the scalar autocorrelation function $\tilde{c}(\tau)$ for three different real-world labelled temporal networks (further information in the text). In panels (b), (d), and (f), we show the autocorrelation-like functions $c_X(\tau)$ obtained from the corresponding  unlabelled network trajectories. 
Panels (a) and (b) are for the US domestic flight network for $\tau$ up to 55 hours with a time resolution of 1 hour; we can observe the daily periodicity. Panels (c) and (d) are for face-to-face contacts in the Malawi village, we show autocorrelation functions up to $\tau=1$ hour, with a time resolution of 1 minute. Panels (e) and (f) show face-to-face contacts at the SFHH conference. The autocorrelation functions are shown up to $ \tau=30$ minutes with a time resolution of 20 seconds. For the Malawi village and the conference, the autocorrelation-like functions $c_X(\tau)$ show a slow decay, highlighting the long-range correlations characterising human contacts \cite{lacasa2022correlations}.}}
\label{fig:real_ACF}
\end{figure}

\medskip \noindent 
Results are summarised in Fig. \ref{fig:real_ACF}. Panels (a) and (b) depict $\tilde{c}({\tau})$ and $c_X(\tau)$ for the labelled and unlabelled US air traffic network trajectory, respectively. These show that the $c_X$ can accurately detect the marked 24 hour periodicity, resulting from the daily flight activity pattern. In the case of face-to-face contact in the Malawi village and the SFHH conference, the autocorrelation of the labelled temporal network exhibits a slow decay [panels (c) and (e)], which is shown to be slightly slower when measured in terms of $c_X(\tau)$ [panels (d) and (f)].

\section{Conclusion}\label{Sec:conclusion}

In this work, we have shown that it is partially possible to retrieve the dynamical fingerprints of a labeled temporal network, even when information on the node labels has been removed. To do this we make use of graph invariants --properties of graphs which are invariant under node relabelling-- to enable the extensions of concepts such as the network autocorrelation function or dynamic instability to unlabelled temporal networks.
We validate our approach by constructing different generative models of temporal networks with and without labels, including models of periodic, correlated and chaotic dynamics. 
Our results demonstrate that chaos in the dynamics of a network can be qualitatively detected even when node labels are removed (Figs.~\ref{fig:Low_chaos} and \ref{fig:High_chaos}).
However, we also find that the pseudo-distances that one can efficiently define for unlabelled network trajectories only grow sub-linearly with respect the true distance between the underlying labelled graphs [Fig.~\ref{fig:small_perturbation}~(b)] and that this sub-linear relation is not universal. This makes it difficult to estimate bona fide Lyapunov exponents from unlabelled network trajectories. On the other hand, we were successful in quantifying (noisy) periodicity and linear temporal correlations in unlabelled network trajectories. 
Overall, this proof of concept provides a promising avenue to characterizing the temporal structure on network trajectories even when information on nodes is absent.

\medskip \noindent 
In broader terms, our work explores the relation between labelled and unlabelled networks, and provides steps towards characterizing the dynamics of graphs without the need to observe time-varying adjacency matrices. Further work should extend our ideas to other graph invariants. One could also define pseudo-distances based not on single graph invariants, but on combinations of them. This would allow one to include information about different aspects of the unlabelled temporal networks. It would then further be interesting to ask what combination of invariants might be best suited to characterise a given type of graph dynamics. 

\medskip \noindent 
Other questions for future work include the extension of these ideas to higher-order network structures such as hypergraphs \cite{gallo2024higher}. Finally, we hope this work will catalyze experimental analysis of unlabelled network trajectories in fields ranging from animal migration, flocking, and pedestrian dynamics to active matter in physics and biology. Progress in this direction would reduce the need for sophisticated tracking mechanisms in these applications.

\subsubsection*{Appendix: Analytical calculation of $d_\text{lab}$ in Eq.~(\ref{Eq:d_ED_in_perturbations})}\label{App:ED_small_perturbations}
 The distance $d_\text{lab}$ in Eq.~(\ref{eq:lab}) counts the number of egdes that exist either only in $G$ or only in $G'$, but not in the respective other graph. Since all edges in $G$ and $G'$ are constructed independently, we can restrict the discussion to one focal edge. If $Z$ is chosen as the total number of node-pairs, $N(N-1)/2$, the average distance $d_{\text{lab}}(G,G')$ reduces to
\begin{equation}
    \avg{d_{\text{lab}}(G,G')}= \mbox{Prob}(x+\xi > p \mid x \leq p) + \mbox{Prob}(x+\xi \leq p \mid x> p),
\end{equation}
where $x$ is a random number drawn uniformly from the interval between zero and one, and where $\xi$ is a Gaussian random number of mean zero and with variance $\sigma^2$. The angle bracket denotes an average over the graphs $G$ and $G'$, in the context of the focal edge this means the average over $x$ and $\xi$. The quantity $  \avg{d_{\text{lab}}(G,G')}$ is thus the probability that the focal edge exists in one graph but not in the other.

\medskip \noindent
We write $x'=x+\xi$ and start from the joint probability density $ P(x', x) $. Given that $x$ is uniformly distributed in $[0,1]$ we have
\[
P(x', x) = \begin{cases}
\frac{1}{\sqrt{2\pi\sigma^2}} \exp\left(-\frac{(x' - x)^2}{2\sigma^2}\right) & \text{if } 0 \leq x \leq 1, \\
0 & \text{otherwise}.
\end{cases}
\]
To find $ P(x' > p \mid x \leq p) $, we integrate the joint pdf:

\[
\mbox{Prob}(x' > p \mid x \leq p) = \int_{p}^\infty dx' \int_{0}^{p}  dx\, P(x', x) \,
\]
Similarly, we have
\[
\mbox{Prob}P(x' \leq p \mid x > p) = \int_{-\infty}^{p} dx' \int_{p}^{1} dx\, P(x', x).
\]
These integrals can all be performed using the error function,
\[
\text{erf}(x) = \frac{2}{\sqrt{\pi}} \int_0^x \exp(-t^2) \, dt
\]
and its integral,
\[
\int dx\, \text{erf}(x) = x\,\text{erf}(x)+ \frac{e^{-x^2}}{\pi}+\mbox{const.}
\]
We then obtain
\begin{eqnarray}
&&\mbox{Prob}(x' > p \mid x \leq p) + \mbox{Prob}(x'\leq p \mid x> p)\nonumber \\
&=&\sigma \sqrt{\frac{2}{\pi}}+\frac{1}{2}-\frac{\sigma}{\sqrt{2 \pi}} \left[\exp\left(-\left(\frac{p}{\sqrt{2} \, \sigma}\right)^2\right) + \exp\left(-\left(\frac{p - 1}{\sqrt{2} \sigma}\right)^2\right)\right]\nonumber \\
&&- \frac{1}{2} \left[p \cdot \text{erf}\left(\frac{p}{\sqrt{2} \, \sigma}\right) + (p - 1) \cdot \text{erf}\left(\frac{p - 1}{\sqrt{2} \, \sigma}\right)\right].
\end{eqnarray}

\subsection*{Acknowledgments} We wish to thank Lluís Arola-Fernández for insightful comments. 
AC acknowledges funding by the Maria de Maeztu Programme (MDM-2017-0711) and the AEI (MICIU/AEI/10.13039/501100011033) under the FPI programme. 
LL acknowledges partial financial support from via project MISLAND (PID2020-114324GB-C22) funded by MICIU/AEI/10.13039/501100011033. Financial support has been received from the Agencia Estatal de Investigación (AEI, MCI, Spain) MCIN/AEI/10.13039/501100011033 and Fondo Europeo de Desarrollo Regional (FEDER, UE) under Project APASOS (PID2021-122256NB-C21/C22) and the María de Maeztu Program for units of Excellence in R\&D, grant CEX2021-001164-M.

\subsection*{Data and code availability} Data and codes will be available upon publication.

\bibliography{sn-bibliography}% 
\end{document}